\documentclass{LMCS}

\usepackage{amssymb,hyperref}
\newcommand{\R}{{\Bbb R}}
\newcommand{\N}{{\Bbb N}}

\newcommand{\Q}{{\Bbb Q}}
\newcommand{\Z}{{\Bbb Z}}
\newcommand{\rinf}{\rightarrow \infty}

\def\doi{1 (2:4) 2005}
\lmcsheading%
{\doi}
{28}
{}
{}
{Sep.~\phantom{0}1, 2004}
{Sep.~26, 2005}
{}

\begin{document}

\title{Comparing hierarchies of total functionals}
\author[D.~Normann]{Dag Normann}
\address{Department of Mathematics, The University 
of Oslo, P.O. Box 1053, Blindern N-0316 Oslo, Norway}
\email{dnormann@math.uio.no}

\keywords{domains, real numbers, intensional and
extensional representations, typed hierarchies, hereditarily total
functionals}
\subjclass{F.4.1}

\begin{abstract} 
In this paper, we will address a problem raised by Bauer, Escard\'{o}
and Simpson. We define two hierarchies of total, continuous
functionals over the reals based on domain theory, one based on an
``extensional'' representation of the reals and the other on an
``intensional'' representation. The problem is if these two
hierarchies coincide. We will show that this coincidence problem is
equivalent to the statement that the topology on the Kleene-Kreisel
continuous functionals of a fixed type induced by all continuous
functions into the reals is zero-dimensional for each type.

As a tool of independent interest, we will construct topological
embeddings of the Kleene-Kreisel functionals into both the extensional
and the intensional hierarchy at each type. The embeddings will be
hierarchy embeddings as well in the sense that they are the inclusion
maps at type 0 and respect application at higher types.
\end{abstract}

\maketitle

\section{Introduction} There are essentially two ways (with many
dialects) to represent the reals 
as data-objects using domains. One way is to use approximations to the
reals in such a way that when two objects approximate the same real,
they 
are consistent in the domain-theoretical sense. This is called {\em an
extensional approach}. Another way is to represent a real via a
sequence 
of integers, representing some approximating sequence. Two sequences
may 
represent the same real, as in $1.000 \cdots$ and $0.999 \cdots$, but
as 
data-objects they are quite different and will be considered as 
inconsistent pieces of information. This is called {\em the
intensional 
approach}.

 We will consider one example of each kind.

Our example of an extensional representation will be via the algebraic domain of closed rational intervals where the reals are represented by ideals of intervals such that the intersection of the sets in the ideal consists of one point.

Our example of an intensional representation will be the binary negative digit representation, essentially representing a real as an infinite sum $$a + \sum_{i = 1}^\infty b_{i}\cdot 2^{-i}$$ where $a \in \Z$  and each $b_{i}$ is in $\{-1,0,1\}$. This will essentially be an admissible representation of the reals as defined by Weihrauch\cite{Weihrauch}.

With the  intensional representation we may consider the representatives for the reals as the total elements of an algebraic domain in a natural way. Thus in both the extensional and the intensional case we may first construct the typed hierarchy of continuous functionals of finite types in the category of algebraic domains, then isolate the hereditarily total functionals in both hierarchies and finally consider the extensional collapse of both hierarchies, see Sections \ref{sec2} and \ref{sec4} for details.

The problem is if these two hierarchies coincide.
\begin{center} {\em Why is this an interesting problem?}\end{center}
One of the main motivations for considering typed hierarchies
of domains 
in the first place is to use them for denotational interpretations of 
programs in some extension of typed $\lambda$-calculus. When we add a 
base type representing the reals, it is because we want to consider 
programs where reals, or representations of reals, are accepted as
data-objects. If the language at hand accepts the reals themselves as 
data-objects we use an extensional hierarchy, while if it accepts 
representations for reals, e.g. in the form of data-streams, we use an
intensional hierarchy. Typically, $RealPCF$ (Escard\'{o} \cite{Martin}) is naturally interpreted over a typed hierarchy of continuous functionals based on the extensional approach while e.g. the approach to typed computability over the reals taken by Di Gianantonio \cite{Gi1,Gi2,Gi3}, and also by Simpson \cite{alex}, makes use of an intensional approach.

 Since non-termination is an important aspect of computations,
it is 
natural to use algebraic or continuous domains when constructing these
hierarchies. As pointed out in e.g. Plotkin \cite{Plotkin}, an important class of 
programs are those that terminate on every relevant input, and thus
the 
hereditarily total objects in a typed hierarchy will be of a special 
interest. Our question is in essence if the notion of a hereditarily
total 
and continuous functional of finite type over the reals is an absolute
notion, or if it is a notion that depends on our choice of
representations of 
real numbers.

 The precise version of the problem we address  was first
formulated by Bauer, Escard\'{o} and Simpson 
\cite{BES}. They proved that at the  first three levels (types 0 - 2),
the hierarchies 
coincide. Moreover, they showed that if the topology on the
Kleene-Kreisel 
continuous functionals of type 2 is zero-dimensional, then the
hierarchies 
also coincide for types at level 3. They further showed that the 
assumption of zero-dimensionality may be weakened, but that equality
of 
the extensional and intensional hierarchies for types at level 3 will
have 
consequences for the topology of $Ct_\N(2)$.

 Since then, Normann \cite{Dag.Hier} analyzed the intensional
hierarchy 
further, and described a representation of it via domains such that
the 
total objects of each type form  dense subsets of the underlying domains.

Both the hierarchy of continuous total functionals based on the extensional representation of the reals and the one based on the intensional representation can be viewed as natural analogues of the Kleene-Kreisel continuous functionals \cite{KL,KR}, where $\N$ is replaced by $\R$. Our first result, which is of independent interest, is that the typed structure of the Kleene-Kreisel continuous functionals may be continuously embedded into the extensional hierarchy over the reals. We will also  prove a similar result for the intensional hierarchy. These results will be stated more precisely in due course. In this paper, these embeddings will be used as tools in analyzing the coincidence problem.

The typed hierarchies in question are obtained as quotient spaces of hereditarily total objects under the relation of being equivalent. In the extensional case, equivalence will be the same as consistency for hereditarily total objects.

We will make use of the  approach from Normann \cite{Dag.Hier}.  One consequence of the density theorem in \cite{Dag.Hier} is that consistency will be an equivalence relation on the hereditarily total objects, and that consistent total objects will be equivalent in the sense of representing the same functional.  We will study the intermediate hierarchy of quotient spaces dividing the hereditarily total objects of our alternative hierarchy just by consistency instead of full equivalence, and see that the topology of these intermediate spaces share the relevant properties  of the Kleene-Kreisel functionals, via mutual topological embeddings. This intermediate hierarchy, the embedding results, the density theorems referred to above and an approximation lemma proved in Normann \cite{Dag.Def} are used together with a technique from Bauer, Escard\'{o} and Simpson \cite{BES}  in order to link the coincidence problem to a problem about the topology of the Kleene-Kreisel functionals. This will be made more precise later.
\subsection*{Organization}
In Section \ref{sec2} we will give the construction of the Kleene-Kreisel continuous functionals, the $Ct_{\N}(k)$ hierarchy, and of the analogue hierarchy of extensional functionals over the reals, the $Ct_{\R}^{E}(k)$-hierarchy. We will also state, and to some extent prove, the relevant properties of these hierarchies and results from topology in general that we will need. There is hardly any original material in this section.

In Section \ref{sec3} we will state and prove the embeddability of the $Ct_{\N}(k)$-hierarchy into the $Ct_{\R}^{E}$-hierarchy.

In Section \ref{sec4} we will introduce the intensional $Ct_{\R}^{I}$-hierarchy and the {\em smoothened version}, the equivalent $Ct_{\R}^{S}$-hierarchy, and we will prove a conditional coincidence theorem.

In Section \ref{sec5} we give a full characterization of the coincidence problem by proving the converse of the main theorem in Section \ref{sec4}.

In Section \ref{sec6} we discuss some further problems, and
in the appendix (Section \ref{sec7}) we will prove a special case of the approximation lemma from Normann \cite{Dag.Def}.
\subsection*{Acknowledgements}
Two anonymous referees of a first version of this paper gave valuable comments.
\section{Background} \label{sec2}We will assume that the reader is familiar
with the theory of Scott Domains, or bounded complete algebraic domains. In this paper all domains in question will be algebraic and bounded complete, i.e. each bounded set will have a least upper bound. These properties will not necessarily be repeated when assumed. We recommend Stoltenberg-Hansen \& al. \cite{viggo}, Abramsky
and Jung
\cite{AJ} or Gierz \& al. \cite{Gierz} for an introduction to domain
theory. For an introduction to the domain theoretical approach to the Kleene-Kreisel continuous functionals \cite{KL,KR} we suggest the handbook paper Normann \cite{Dag.Hand}.

 We will describe the construction of the Kleene-Kreisel  continuous
functionals and the corresponding hierarchy over the reals based on
the  extensional representation mainly by setting the notation to be
used in the  paper.

 We will restrict our attention to the pure types. In this paper we will let $\N$ denote the non-negative integers.
\begin{defi}{Let $N(0) = \N_\bot$, with $\bar N(0) = \N$ and
$n \approx^N_0m$ if and 
only if $n = m \in \N$. Let $N(k+1) = N(k) \rightarrow N(0)$ in the
category of algebraic domains. 
For $f,g \in N(k+1)$, let $$
f \approx^N_{k+1} g \Leftrightarrow \forall a \in N(k) \forall b \in 
N(k)(a \approx^N_k b \Rightarrow f(a) \approx^N_0 f(b)).$$
Let $\bar N(k+1) = \{f \in N(k+1)\;;\; f 
\approx^N_{k+1}f\}$.}
\end{defi}
$\approx^N_{k}$ will be a partial equivalence relation. The elements of $\bar N_{k}$ will be called the {\em hereditarily total functionals}, and equivalence will mean that equivalent functionals will give us the same well defined output to a hereditarily total input.

Since $\approx^N_{k}$ is both symmetric and transitive, we will have that if $x$ and $y$ are elements of $N(k)$ such that $x \approx^N_{k} y$, then $x \in \bar N(k)$.
\begin{prop} For $x,y \in N(k)$, we have that $x \approx^N_k y
\Leftrightarrow x \sqcap y 
\in \bar N(k)$. 

 For $x,y \in \bar N(k)$ we have that $x 
\approx^N_k y \Leftrightarrow x$ and $y$ are consistent.
\end{prop} For a proof, see e.g. Normann \cite{Dag.Hand}. The
first part of this 
proposition was originally proved in Longo and Moggi \cite{LM}, and
the 
second part is a consequence of the domain-theoretical version of the Kleene-Kreisel Density Theorem, see Proposition \ref{pr3}.
\begin{defi}\label{def2}{By recursion on $k$ we define the set $Ct_{\N}(k)$ and the projection map
$\rho^N_k:\bar N(k)\rightarrow Ct_{\N}(k)$ as 
follows:
\begin{itemize}
\item[] $\rho^N_0(n) = n$, $Ct_{\N}(0) = \N$.
\item[] As an induction hypothesis, an arbitrary element of $Ct_{\N}(k)$ will be of the form $\rho_{k}^N(x)$ where $x \in {\bar N}(k)$. If $f \in \bar N(k+1)$, we let 
$$\rho^N_{k+1}(f)(\rho^N_k(x)) = f(x).$$ This is well defined by the definition of $\approx^N_{k+1}$, assuming that $\rho^N_{k}$ identifies exactly $\approx^N_{k}$-equivalent objects, and then $\rho^N_{k+1}$ will identify exactly $\approx^N_{k+1}$-equivalent objects.
\item[] Let $Ct_\N(k+1) = \{\rho^N_{k+1}(f)\;;\;f \in \bar N(k+1)\}$.
\end{itemize} $Ct_\N(k)$ is known as the {\em Kleene-Kreisel continuous
functionals\/} of 
type $k$.}
\end{defi}
The {\em topology} on $Ct_{\N}(k)$ will be the finest topology such that $\rho^\N_{k}$ is continuous. Then $Ct_{\N}(k+1)$ will consist of exactly all continuous maps $F:Ct_{\N}(k) \rightarrow \N$.

 Using similar 
constructions, we will now define a hierarchy of functionals over the
reals. This will be based on the extensional representation of 
reals, and we will use the letter $E$ for {\em extensional} to denote this hierarchy.
\begin{defi}{Let $E(0)$ be the algebraic domain of ideals
over $$
\{\R\} \cup \{[p,q]\;;\;p \in \Q \wedge q \in \Q \wedge p \leq q\}$$ where the intervals are
ordered by reverse inclusion.

 Let $E(k+1) = E(k) \rightarrow E(0)$.

 If $\alpha$ is an ideal in $E(0)$, then $\cap \alpha \neq
\emptyset$. We let
\begin{center}$\alpha \approx^E_0 \beta \Leftrightarrow \cap \alpha =
\cap 
\beta = \{x\}$ for some $x \in \R$.
\end{center} We define $\approx^E_k$ by recursion on $k$ in analogy
with the definition 
of $\approx^N_k$, and let $\bar E(k) = \{x \in E(k)\;;\;x \approx^E_k 
x\}$.}
\end{defi} We then have
\begin{prop}{\em(Normann \cite{Dag.Sh})}

 For $x,y \in E(k)$ we have that $x \approx^E_k y
\Leftrightarrow x \sqcap 
y \in \bar E(k)$.

 For $x,y \in \bar E(k)$ we have that $x \approx^E_k y
\Leftrightarrow x$ 
and $y$ are consistent.
\end{prop} 
\begin{defi}{We define $Ct_{\R}^{E}(k)$ and $\rho_{k}^{E}: \bar E(k) \rightarrow Ct_{\R}^{E}$ in analogy with Definition \ref{def2} as follows:
We let $\rho^E_0(\alpha) = x$ if $\{x\} = \cap
\alpha$.

 We let $\rho^E_{k+1}(x)(\rho^E_k(y)) =
\rho^E_0((x)(y))$.

 We let  $Ct^E_\R(k) = \{\rho^E_k(x)\;;\;x \in \bar E(k)\}$.
}\end{defi} The domains $E(k)$ are special instances of
domains $E(\sigma)$ for all 
types $\sigma$. Then the $E$-hierarchy may be used to implement 
Escard\'{o}'s   $Real PCF$ \cite{Martin}, though the
approach via continuous  domains is the one used originally. 

We define the topology on $Ct_{\R}^{E}(k)$ as the finest topology making $\rho_{E}(k)$ continuous. Then $Ct_{\R}^{E}(k+1)$ is exactly the set of continuous functions from $Ct_{\R}(k)$ to $\R$.
\begin{prop}\label{pr3}\hfill
\begin{itemize}
\item[{a)}]For each $k \in \N$, $\bar N(k)$ is
dense in $N(k)$.
\item[{b)}]For each $k \in \N$, $\bar E(k)$ is dense in $E(k)$. \end{itemize}
\end{prop}
Part a) is the domain-theoretical version of the Kleene-Kreisel Density Theorem. Part b) is proved in \cite{Dag.Sh}

In this paper we will work with $\omega$- algebraic domains $A$ ( meaning that the set of compacts is countable), a set $\bar A \subseteq A$ of  ``total'' objects, an equivalence relation $\approx_{A}$ on $\bar A$ and (essentially) the quotient topology on $\bar A/\approx_{A}$.

This topology will be {\em sequential}, which means that it is generated from the set of convergent sequences with limits.

Since $A$ is $\omega$-algebraic, we see that any open covering of a subset of $\bar A /\approx_{A}$ can be replaced by a countable sub-covering, i.e. the spaces are {\em hereditarily Lindel\"{o}f}.

These are facts of basic general topology.

In the sequel we will also make use of the following  fact:
\begin{lem}\label{lem.gen}
Let $X$ be a topological space, $\rho:X \rightarrow Y$ an onto map and let $Y$ be equipped with the identification topology, i.e. $O \subseteq Y$ is open if and only if $\rho^{-1}[O]$ is open in $X$.

Let $Z \subseteq Y$ be closed, let $Z_{1} = \rho^{-1}[Z]$, and let $\rho_{Z}$ be $\rho$ restricted to $Z_{1}$.

Then the identification topology on $Z$ induced from $\rho_{Z}$ and the subspace topology on $Z_{1}$ will coincide with the subspace topology on $Z$.\end{lem}
The proof is elementary and is left for the reader.

Following Scarpellini \cite{scarp}  on one hand (see also \cite{hyland} or \cite{Dag.Hand}) and Normann \cite{Dag.Sh} on the other, we also have that the hierarchies $\{Ct_{\N}(k)\}_{k \in \N}$ and $\{Ct_{\R}^{E}(k)\}_{k \in \N}$ can be defined in the category of Kuratowski limit spaces, see Kuratowski \cite{Kuratowski}. We have the canonical limit structures on $\N$ and $\R$ resp. Then, by recursion on $k$ we may define  $Ct_{\N}(k)$ and $Ct_{\R}^{E}(k)$ with limit structures as follows, where $Ct(k)$ may stand for both sets:

$F \in Ct(k+1)$ if $F:Ct(k) \rightarrow Ct(0)$ and for all $\{a_{n}\}_{{n \in \N}}$ and $a$ from $Ct(k)$, $a = \lim_{n \rightarrow \infty}a_{n}  \Rightarrow F(a) = \lim_{n \rightarrow \infty} F(a_{n})$,
i.e., $F$ is sequence continuous.

$F = \lim_{n \rightarrow \infty}F_{n}$ if for all $\{a_{n}\}_{{n \in \N}}$ and $a$ from $Ct(k)$, 
 $a = \lim_{n \rightarrow \infty}a_{n}  \Rightarrow F(a) = \lim_{n
 \rightarrow \infty} F_{n}(a_{n})$.

For both hierarchies of quotient spaces it is also the case that the convergent sequences of quotients with limits are exactly the sequences obtained by taking a convergent sequence at the domain level and then take the quotients.

For the Kleene-Kreisel continuous functionals there is a third characterization of the convergent sequences as well. The following proposition summarizes all this. Part a)  was proved in Hyland \cite{hyland}, for an exposition see \cite{Dag.Hand}. Part b) was proved in Normann \cite{Dag.Sh}.
\begin{prop}\label{prop.lim}\hfill \begin{itemize}
\item[{a)}]Let $k \in \N$, $\{f_{i}\}_{i \in \N}$ be a sequence from $Ct_{\N}(k+1)$ and $f \in Ct_{\N}(k+1)$. Then the following are equivalent:\begin{itemize}
\item[{i)}]$f = \lim_{i \rightarrow \infty}f_{i}$ in the topology on $Ct_{\N}(k+1)$.
\item[{ii)}]Whenever $a = \lim_{i \rightarrow \infty}a_{i}$ in $Ct_{\N}(k)$, then $f(a) = \lim_{i \rightarrow \infty}f_{i}(a_{i})$ in $\N$.
\item[{iii)}]There is a convergent sequence with limit $g = \lim_{i \rightarrow \infty} g_{i}$ from $\bar N(k+1)$ such that $f = \rho^N_{k+1}(g)$ and $f_{i} = \rho^N_{k+1}(g_{i})$ for each $i \in \N$
\item[{vi)}]There is a continuous {\em modulus} $g$ {\em of convergence}, i.e. $g \in Ct_{\N}(k+1)$ such that $$\forall a \in Ct_{\N}(k) \forall i \geq g(a) (f_{i}(a) = f(a)).$$\end{itemize}
\item[{b)}]Let $k \in \N$, $\{f_{i}\}_{i \in \N}$ be a sequence from $Ct_{\R}^{E}(k+1)$ and $f \in Ct_{\R}^{E}(k+1)$. Then the following are equivalent:\begin{itemize}
\item[{i)}]$f = \lim_{i \rightarrow \infty}f_{i}$ in the topology on $Ct_{\R}^{E}(k+1)$.
\item[{ii)}]Whenever $a = \lim_{i \rightarrow \infty} a_{i}$ in $Ct_{\R}^{E}(k)$, then $f(a) = \lim_{i \rightarrow \infty}f_{i}(a_{i})$ in $\R$.
\item[{iii)}]There is a convergent sequence with limit $g = \lim_{i \rightarrow \infty}g_{i}$ from $\bar E(k+1)$ such that $f = \rho^E_{k+1}(g)$ and $f_{i} = \rho^E_{k+1}(g_{i})$ for each $i \in \N$.\end{itemize}
\end{itemize}\end{prop}
We will not prove these results in detail, but, as we will see, they follow from the next proposition and a similar proposition for the reals:
\begin{prop}\label{proposition.new}
Let $(X,\sqsubseteq)$ be an $\omega$-algebraic domain and let $\bar X \subseteq X$ be a subset that is uppwards closed.
If $f:\bar X \rightarrow \N$ is continuous in the subspace topology on $\bar X$, then $f$ can be extended to a continuous $g:X \rightarrow \N_{\bot}$. \end{prop}
Proposition \ref{proposition.new} is due to the author, but the proof was not published. If we replace $\N$ by $\R$, the proposition is proved as Lemma 6.2 in Normann \cite{Dag.Sh}, and the proof from \cite{Dag.Sh} works in this simpler case as well. In the present paper, the method of proof is employed  in the proof of Theorem \ref{th.apprx}, in defining $Y^f_{n}$ from $X^f_{n}$. The argument is also used in Normann \cite{Dag.Def}.
\begin{cor}\label{corollary.new} In Proposition \ref{proposition.new} we may replace $\N$ with $\N^\N$ and $\N_{\bot}$ with $\N_{\bot} \rightarrow \N_{\bot}$.\end{cor}

\proof
Let $g:\bar X \rightarrow \N^\N$ be continuous.

Let $g_{1}(x,n) = g(x)(n)$ and apply Proposition \ref{proposition.new} to $g_{1}:\bar X \times \N \rightarrow \N$.\qed

In this paper, function spaces, and subspaces of function spaces, will play an important part. There is no canonical topology on a function space, so we will now discuss the topology we will use for the relevant cases.

For the purpose of this discussion, let $(X,\bar X , \approx_{X})$ be an $\omega$-algebraic domain $X$ with a totality $\bar X$ satisfying the requirement of Proposition \ref{proposition.new} and let $\approx_{X}$ be an equivalence relation on $\bar X$. Let ${\mathcal T}(X)$ be $\bar X / \approx_{X}$ with the quotient topology. Let ${\mathcal T}(X) \rightarrow \N$ be the set of continuous functions from ${\mathcal T}(X)$ to $\N$. We will define a ``default'' topology on ${\mathcal T}(X) \rightarrow \N$ and establish some properties of this topology.
These properties will extend to ${\mathcal T}(X) \rightarrow \N^\N$, and then to ${\mathcal T}(X) \rightarrow \Z \times \{-1,0,1\}^{\N^+} $, for which we will use them. ($\N^+$ is the set of positive integers.)

Let $f \in {\mathcal T}(X) \rightarrow \N$. By Proposition \ref{proposition.new} there is a continuous $g:X \rightarrow \N_{\bot}$ such that for each $a \in {\mathcal T}(X)$ and each $x \in a$ we have that $g(x) = f(a)$.

Let $Y = X \rightarrow \N_{\bot}$ in the category of algebraic domains. Let $g\in Y$ and $h \in Y$. Let
$$g \approx_{Y} h \Leftrightarrow \forall x \in \bar X \forall y \in \bar X (x \approx_{X} y \rightarrow g(x) = h(y) \in \N).$$
Then $\approx_{Y}$ will be a partial equivalence relation, and we let $$\bar Y = \{g \in Y\;;\; g \approx_{Y} g\}.$$

We use the quotient topology on $\bar Y / \approx_{Y}$ and the canonical 1-1 correspondence between $\bar Y / \approx_{Y}$ and ${\mathcal T}(X) \rightarrow \N$ to define the topology on ${\mathcal T}(X) \rightarrow \N$.

\begin{defi}{Let $(X,\bar X , \approx_{X})$ be as above.

We say that ${\mathcal T}(X)$ {\em accepts liftings of convergent sequences} if whenever $a = \lim_{n \rinf} a_{n}$ in ${\mathcal T}(X)$ then there is an $x \in a$ and an $x_{n} \in a_{n}$  for each $n$ such that $x = \lim_{n \rinf} x_{n}$.}\end{defi}
\begin{lem}\label{lemma.210405}
Let $(X,\bar X , \approx_{X})$ and ${\mathcal T}(X)$ be as above and assume that ${\mathcal T}(X)$ accepts liftings of convergent sequences. Let $(Y, \bar Y , \approx_{Y})$ and ${\mathcal T}(X) \rightarrow \N$ be as constructed.
\begin{itemize}
\item[a)] ${\mathcal T}(Y)$ accepts liftings of convergent sequences.
\item[b)] If $f \in {\mathcal T}(X) \rightarrow \N$ and $f_{n} \in{\mathcal T}(X) \rightarrow \N$ for each $n \in \N$, then the following are equivalent:
\begin{itemize}
\item[i)] $f = \lim_{n \rinf} f_{n}$.
\item[ii)] Whenever $a = \lim_{n \rinf} a_{n}$ in ${\mathcal T}(X)$, then
$f(a) = \lim_{n\rinf}f_{n}(a_{n}).$\end{itemize}\end{itemize}\end{lem}

\proof
Let $\Omega$ be the domain with compacts $\bot$ and $n$ and $n*$ for $n \in \N$ ($n*$ is just a formal object), where $\bot \sqsubseteq a$ for all compacts $a$, $n \sqsubseteq m$ if and only if $n = m$, $n* \sqsubseteq m*$ if and only if $n \leq m$ and $n* \sqsubseteq k$ if and only if $n \leq k$. $\Omega$ is known as the ``lazy natural numbers''.

We will let $\omega$ be the maximal ideal generated from $\{n*\;;\;n \in \N\}$, and notation-wise we will also use ``$n$'' for the ideal generated by $\{n\}$. Then $\omega = \lim_{n \rinf}n$. Let $\bar \Omega = \N \cup\{\omega\}$.

The point is that a convergent sequence $f_{\omega} = \lim_{ n \rinf} f_{n}$ of continuous functions from ${\mathcal T}(X)$ to $\N$ will be the continuous image of $n \mapsto f_{n}$ where $n \in \bar \Omega$.
\begin{itemize}
\item[a)] Let $f_{\omega} = \lim_{n \rinf }f$. Let $F(n,x) = f_{n(x)}$ for $n \in \bar \Omega$.

\noindent By Proposition \ref{proposition.new} there is a continuous $G:\Omega \times X \rightarrow \N$ such that $G(n,x) = f_{n}([x])$ (where $[x]$ is the equivalence class of $x$) for all $n \in \bar \Omega$ and all $x \in \bar X$.

\noindent Let $g_{n}(x) = G(n,x)$. Then $g_{\omega} = \lim_{n \rinf} g_{n}$. This will be a lifting of the convergent sequence.
\item[b)] Let $f_{\omega} = \lim_{n \rinf}f_{n}$ and let $a_{\omega} = \lim_{n \rinf} a_{n}$.

\noindent By the assumption that ${\mathcal T}(X)$ accepts liftings of convergent sequences and by a) of this lemma, it follows that $f(a) = \lim_{n \rinf}f_{n}(a_{n})$.

\noindent Assume now that $f_{\omega}(a_{\omega}) = \lim_{n \rinf}f_{n}(a_{n})$ whenever $a_{\omega} = \lim_{n \rinf} a_{n}$.

\noindent This means that $$F(n,a) = f_{n}(a)$$
is sequence-continuous on $\bar \Omega \times {\mathcal T}(X)$ (the sequence topology may be finer than the product topology).

\noindent Let $Z$ be the domain $\Omega \times X$, $\bar Z = \{(n,x)\;;\;x \in \bar X \wedge n \in \bar \Omega\}$ and let $(n,x) \approx_{Z} (m,y) \Leftrightarrow n = m \wedge x \approx_{X} y.$
Then the quotient topology on $\bar Z / \approx_{Z}$ is sequential, and is essentially the same as the sequence topology on $\bar \Omega \times {\mathcal T}(X)$.

\noindent Then, by Proposition \ref{proposition.new} there is a continuous $G:\Omega \times X \rightarrow \N_{\bot}$ such that whenever $a \in {\mathcal T}(X)$, $ x \in a$ and $n \in \bar \Omega$ we have that
$$G(n,x) = F(n,a) = f_{n}(a).$$
Let $g_{n}(x) = G(n,x)$ for $n \in \bar \Omega$ and $x \in X$. Then $g_{\omega} = \lim_{n \rinf}g_{n}$ and we may conclude that $f_{\omega} = \lim_{n \rinf}f_{n}$.\qed
\end{itemize}

\begin{rem}\label{remark.revision}{As pointed out, a convergent sequence $f_{\omega} = \lim_{n \rightarrow \infty} f_{n}$ corresponds to a continuous map $f:\bar \Omega \rightarrow ({\mathcal T}(X) \rightarrow \N)$

By the standard isomorphism, $f$ corresponds to a map $\hat f : \bar \Omega \times {\mathcal T}(X) \rightarrow \N$.

$\hat f$ needs not be continuous in the product topology, but when ${\mathcal T}(X)$ accepts liftings of convergent sequences, $\hat f$ is continuous in the quotient topology on $\bar \Omega \times {\mathcal T}(X)$. This will also hold when we replace $\N$ by $\R$.}\end{rem}
A set $A$ in a topological space  $T$ is called {\em clopen} if $A$ is both closed and open. We will let ${\mathcal Z}( T)$ be the subtopology generated by the clopen sets. If $f \in Ct_{\N}(k+1)$ for some $k$, $f$ will be continuous with respect to ${\mathcal Z}(Ct_{\N}(k))$.
\begin{defi}{Let $A \subseteq Ct_\N(k)$. 

 $A$ is a
$\tilde \Pi^0_1$-set if there are clopen sets $A_n$ such  that $A
=
\bigcap_{n \in \N} A_n$.}
\end{defi} 
In this case the $\tilde \Pi^0_{n}$-sets are exactly the sets closed in ${\mathcal Z}(Ct_{\N}(k))$.
When $A$ is a $\tilde \Pi^0_1$-set in $Ct_\N(k)$, then the topology on
$A$ induced from the topology on $Ct_\N(k)$ will coincide with the
quotient topology induced by $$
\bar A = \{x \in \bar N(k)\;;\;\rho^\N_k(x) \in A\}$$
and $\approx^\N_k$, see Lemma \ref{lem.gen}.

If $A \subseteq Ct^{E}_{\R}(k)$ is an arbitrary subset, we will consider the quotient topology on $A$ induced by $\rho^{E}_{k}$ restricted to $(\rho^{E}_{k})^{-1}(A)$.

 $\{f:A \rightarrow \R\;;\;f$ is continuous$\}$ will have a default topology in analogy with ${\mathcal T}(X) \rightarrow \N$, and by the $\R$-versions of Proposition \ref{proposition.new} and its consequences, this will be the  finest topology where $f = \lim_{n \rinf} f_{n}$ exactly when
$$\forall a \in A \forall \{a_{n}\}_{n \in \N} \in A^{\N}( a = \lim_{n \rinf}a_{n} \Rightarrow f(a) = \lim_{n \rinf}f_{n}(a_{n})).$$
We use this topology in Proposition \ref{approx.lemma} a) below.
 Part a) is the Approximation Lemma, i.e. Theorem 2, from Normann \cite{Dag.Def}. Part b) is essentially a special case of a), and will be proved in the Appendix for the sake of completeness.
\begin{prop}\label{approx.lemma}\hfill
\begin{itemize}
\item[{a)}]Let $A \subseteq Ct^{E}_\R(k)$ and let $f:A \rightarrow
\R$ be continuous. Then, continuously in $f$, there are $f_n \in
Ct^{E}_\R(k+1)$ such that whenever $x \in A$ and $x = \lim_{n \rinf} x_n$
with each $x_n \in Ct^{E}_\R(k)$ we have that $f(x) =
\lim_{n \rinf}f_n(x_n)$.
\item[{b)}]Let $A \subseteq Ct_\N(k)$ and let $f:A \rightarrow
\N$ be continuous. Then, continuously in $f$, there are $f_n \in
Ct_\N(k+1)$ such that whenever $x \in A$ and $x = \lim_{n \rinf} x_n$
with each $x_n \in Ct_\N(k)$ we have that $f(x) =
\lim_{n \rinf}f_n(x_n)$.
\end{itemize}\end{prop}

\section{A hierarchy  of embeddings}\label{sec3} 
\subsection{Aim and consequences}
$N(k)$ and $E(k)$ are examples of domains with totalities $\bar N(k)$ and $\bar E(k)$ resp. A continuous map $\pi:N(k) \rightarrow E(k)$ is then called {\em total} if $\pi$ maps $\bar N(k)$ into $\bar E(k)$.

We are operating with equivalence relations $\approx_{k}^{N}$ and $\approx_{k}^{E}$ on $\bar N(k)$ and $\bar E(k)$ coinciding with consistency, and then a continuous total map $\pi:N(k) \rightarrow E(k)$ will induce a continuous map $\bar \pi:Ct_{\N}(k) \rightarrow Ct_{\R}^{E}(k)$.

In this section we will prove the
following
\begin{thm}\label{Th.emb} For each $k \in \N$ there is a total,
continuous map 
$$\pi_k:N(k) \rightarrow E(k)$$ such that\begin{itemize}
\item[{i)}]$\pi_{0}:N(0) \rightarrow E(0)$ sends a number $n$ to its representative in $E(0)$.
\item[{ii)}]For each $k \in \N$,  for each $f \in \bar N(k+1)$
and 
$a \in \bar N(k)$ we have $$
\pi_{0}(f(a)) = \pi_{k+1}(f)(\pi_k(a)).$$\end{itemize}
\end{thm} 
Before entering the proof of the theorem, we will establish some consequences.

\begin{cor} \label{cor.2}For each $k \in \N$ there is an injective, continuous map
$$
\bar
\pi_k:Ct_\N(k) \rightarrow Ct^E_\R(k)$$
such that \begin{itemize}
\item[{i)}]$\bar \pi_{0}$ is the standard inclusion map from $\N$ to $\R$.
\item[{ii)}]For each $k \in \N$ , each  $F \in Ct_\N(k+1)$ and each $a \in Ct_{\N}(k)$ we have that  $F(a) = \bar \pi_{k+1}(F)(\bar \pi_k(a))$.\end{itemize}
\end{cor} 
\begin{rem}{Independently, Bauer and Simpson \cite{BS} gave a
proof 
for Corollary \ref{cor.2} for $k \leq 2$. Their result is stronger in
the 
sense that it is proved in intuitionistic logic for constructive 
analysis.}
\end{rem}
Another important consequence is the following
\begin{cor} \label{cor.180405}Let $\bar \pi_{k}$ be obtained from Theorem \ref{Th.emb} as in Corollary \ref{cor.2}.

Then the range of $\bar \pi_{k}$ is a closed subset of $Ct_{\R}^{E}(k)$ homeomorphic to $Ct_{\N}(k)$. \end{cor}
\proof

By recursion on $k$ we will define continuous, partial inverses $\pi_{k}^{-1}:E(k) \rightarrow N(k)$ as follows:
\begin{itemize}
\item If $n \in \N$ and $[p,q] \subseteq (n - \frac{1}{3},n + \frac{1}{3})$, we let $\pi_{0}^{-1}([p,q]) = n$, while $\pi_{0}^{-1}([p,q]) = \bot_{\N}$ if the above rule does not apply.
\item If $g \in E(k+1)$ and $a \in N(k)$ we let 
$$\pi_{{k+1}}^{-1}(g)(a) = \pi_{0}^{-1}(g(\pi_{k}(a))).$$
\end{itemize}
By induction, i) and ii) below  follow from the construction:
\begin{itemize}
\item[i)] If $k \in \N$ and $a \in N(k)$ then $a = \pi_{k}^{-1}(\pi_{k}(a))$.
\item[ii)] If $a \in \bar E(k)$, $b \in \bar E(k)$, $a \approx_{k}^{E} b$ and $\pi_{k}^{-1}(a) \in \bar N(k)$, then $\pi_{k}^{-1}(b) \approx_{k}^N \pi_{k}^{-1}(a)$ and consequently $\pi_{k}^{-1}(b) \in \bar N(k)$.
\end{itemize}
This shows that $Ct_{\N}(k)$ is homeomorphic to the range of $\bar \pi_{k}$ with the quotient topology. We have to show that this range is closed, and then the rest of the corollary will follow from Lemma \ref{lem.gen}.
In order to prove that the range is closed, we use that the topology is sequential, see Proposition \ref{prop.lim}.

The range of $\bar \pi_{0}$ is just the closed subset $\N$ of $\R$.

If $f \in \bar E(k+1)$ we have that $\pi^{-1}_{k+1}(f) \in \bar N(k+1)$ if
$$\forall a \in \bar N(k)(f(\pi_{k}(a)) \in \N).$$
Let $g = \lim_{i \rightarrow \infty}g_{i}$ in $Ctæ_{\R}(k+1)$ such that each $g_{i}$ is in the range of $\bar \pi_{k+1}$. We will show that the range of $\bar \pi_{k+1}$ is closed by showing that $g$ is in the range of $\bar \pi_{k+1}$.

By Proposition \ref{prop.lim}, let $h = \lim_{i \rightarrow \infty}h_{i}$ where $h \in \bar E(k+1)$, each $h_{i} \in \bar E(k+1)$, $g = \rho^{E}_{k+1}(h)$ and each $g_{i} \in \rho^{E}_{k+1}(h_{i})$.

Then
$\forall a \in \bar N(k)(h(\pi_{k+1}(a)) \in \N)$, so $\pi^{-1}_{k+1}(h)$ is defined. Then
$$\pi_{k+1}(\pi^{-1}_{k+1}(h)) = \lim_{i \rightarrow \infty}\pi_{k+1}(\pi^{-1}_{k+1}(h_{i})) = \lim_{i \rightarrow \infty}h_{i} = h.$$
It follows that $h$ is in the range of $\pi_{k+1}$, so $g$ is in the range of $\bar \pi_{k+1}$, and the corollary is proved.\qed

During the construction we will observe that  $\pi_k$ is computable such that
if $a \in 
N_k$ is not total, then $\pi_k(a)$ is not total. By Kreisel \cite{KR} we know that $\bar N(k+1)$ is complete $\Pi^1_{k}$ for $k \geq 1$, see also Normann \cite{Dag.Hand}. It is strait-forward  to show by induction on $k$ that $\bar E(k+1)$ is a $\Pi^1_{k}$-set for $k \geq 1$. We then obtain
\begin{cor}\label{cor.tot}For $k \geq 1$, $\bar E(k+1)$ is complete $\Pi_k^1$.
\end{cor}
The following result is proved using Proposition \ref{approx.lemma} and Corollary \ref{cor.2}. By a suitable adjustment of Proposition \ref{approx.lemma}, the use of Corollary \ref{cor.2} may be avoided.
\begin{thm}\label{Theorem.ext} Let $A \subseteq Ct_\N(k)$ be a
$\tilde \Pi^0_1$-set. 
\begin{itemize}
\item[{a)}]If $f:A \rightarrow \R$ is
continuous,  then $f$ may be extended to a continuous $g:Ct_\N(k)
\rightarrow \R$ such that the map $f \mapsto g$ is continuous. 
\item[{b)}] If $f:A \rightarrow \N$ is
continuous,  then $f$ may be extended to a continuous $g:Ct_\N(k)
\rightarrow \N$ such that the map $f \mapsto g$ is continuous.
\end{itemize} 
\end{thm} 

\proof
 
Both a) and b) are trivial when $k = 0$ so assume that $k > 
0$.

 We prove a). The proof of b) is similar, but simpler.  

 Let $\bar
\pi_k:Ct_\N(k) \rightarrow Ct^E_\R(k)$ be as in Corollary 
\ref{cor.2}. Let $B$ be the image of $A$ under $\bar \pi_k$. If $x$ is
in 
the range of $\bar \pi_k$ we see from Corollary \ref{cor.2} that
$$
x = \bar \pi_{k}(x \circ \bar \pi_{k-1}).$$
Then $f'$ defined by $f'(x) = f(x \circ \bar \pi_{k-1})$ is a continuous
map 
from $B$ to $\R$, continuously depending on $f$.

 By Proposition \ref{approx.lemma} a) there is
a sequence 
$\{f'_n\}_{n \in \N}$ from $Ct^E_\R(k+1)$ such that whenever $x \in
B$, 
$\{x_n\}_{n \in \N}$ is a sequence from $Ct^E_\R(k)$ and $x = \lim_{n 
\rinf} x_n$, then $f'(x) = \lim_{n \rinf}f'_n(x_n)$.

 Let $f_n(z) = f'_n(\bar \pi_k(z))$.

 Let $A = \bigcap_{n \in \N}A_n$ where each $A_n$ is
clopen.

 Let
\begin{itemize}
\item $g(z) = f(z)$ if $z \in A$.
\item $g(z) = f_n(z)$ for the least $n$ such that $z \not \in A_n$
otherwise.
\end{itemize} Then $g$ is a continuous extension of $f$.

The construction of $g$ from $f$ is by composing continuous operators, so $g$ depends continuously on $f$.

 In order to prove b) we use part b) of Proposition \ref{approx.lemma}
 in a similar way.

 This ends the proof of Theorem \ref{Theorem.ext}.\qed

\begin{cor}\label{cor.3.15} If $A \subseteq Ct_\N(k)$ is
$\tilde \Pi^0_1$, then $A \rightarrow \N$ is 
homeomorphic to a $\tilde \Pi^0_1$-subset of $Ct_\N(k+1)$.
\end{cor} 

\proof

 Let $f:A \rightarrow \N$. By
Theorem \ref{Theorem.ext} there will be an 
extension $f_1:Ct(k) \rightarrow \N$ of $f$, continuous in $f$.
Clearly, 
if $f_1 \in Ct_\N(k+1)$, then the restriction of $f_1$ to $A$ is
continuous 
in $f_1$, so $A\rightarrow \N$ and 
$$
B = \{f_1 \in Ct_\N(k+1)\;;\;f \in A \rightarrow \N\}$$
are homeomorphic.

 Let $\{x_i\}_{i \in \N}$ be a dense subset of $Ct_\N(k)$.
Then $$
g \in B \Leftrightarrow \forall i (g(x_i) = (g \upharpoonright
A)_1(x_i))$$
and this is $\tilde \Pi^0_1$.\qed

\subsection{Some machinery}
\begin{defi}{For each $k \geq 0$ and $a \in Ct_\N(k)$ we 
define the $n$'th approximation $a_n$ to $a$ as follows:

 For $k = 0$ we let $a_n = a$ if $a \leq n$ and $a_n = 0$ if
$n < a$.

 For $k > 0$ we let $a_n(x) = (a(x_n))_n$.}
\end{defi}
\begin{lem} \label{lemma2.2}{\em(Essentially Grilliot \cite{gri})}

 For each $k \in \N$ and $a \in Ct_\N(k)$, we have that $a =
\lim_{n 
\rightarrow \infty} a_n$.
\end{lem} For the sake of completeness, we give the proof. The point
is that 
along with the proof, we give an algorithm for a {\em modulus of 
convergence\/} uniformly in the given $a$, i.e. when $k \geq 1$ we will give the algorithm for a map
$$M_{k}:Ct_{\N}(k-1) \times Ct_{\N} \rightarrow \N$$
such that
$$\forall a \in Ct_{\N}(k)\forall b \in Ct_{\N}(k-1) \forall m \in \N(m \geq M_{k}(b,a) \rightarrow a_{m}(b) = a(b)).$$
By Proposition \ref{prop.lim} the existence of this modulus suffices to prove the lemma. We also define the modulus $M_{0}:\N \rightarrow \N$.

 In this proof we will observe the following conventions:

 $n$, $i$, $j$ etc. will denote natural numbers. $f$, $g$ etc.
will denote 
functions, or functionals one type below the type in question. $F$, 
$G$ etc. will denote functionals of the type $k$ in question, when $k 
\geq 2$. We will use induction on $k$.

\proof

 $k = 0$: Clearly $lim_{n
\rightarrow \infty}i_n = i$ with modulus 
$i$, i.e. $n \geq i \Rightarrow i_n = i$. Thus we let $M_{0}(i) = i$

 $k = 1$: Clearly
$lim_{n \rightarrow \infty} f_n(i) = \lim_{n 
\rightarrow \infty}(f(i_n))_n = f(i)$ with modulus $g(i) = max\{i , 
f(i)\}$. Thus we let $M_{1}(i,f) = max\{i,f(i)\}$.

 $k > 1$: It is sufficient to show that $F(f) = \lim_{n
\rightarrow 
\infty}F(f_n)$ and to compute a modulus $G$ for this.

 Convergency follows from the fact that $F$ is continuous and that $f = \lim_{n \rightarrow \infty}f_{n}$.

 Let $g$ be the modulus for $f = \lim_{n \rightarrow \infty}
f_n$ 
obtained by the induction hypothesis, i.e. $g(\xi) = M_{k-1}(\xi,f)$.

 For each $n$ and $\xi \in Ct_\N(k - 2)$, let
\begin{enumerate}
\item $h_n(\xi) = f_m(\xi)$ for the least $m$ such that $n \leq m <
g(\xi)$ 
and $F(f_m) \neq F(f)$ if there is such $m$.
\item $h_n(\xi) = f_{g(\xi)}(\xi)$ if there is no such $m$.
\end{enumerate} Then $h_n = f$ if there is no $m \geq n$ such that
$F(f_m) \neq 
F(f)$, while $h_n = f_m$ for the least $m \geq n$ with $F(f_m) \neq 
F(f)$ otherwise.

 Let $G(f) = max\{F(f),\mu n(F(h_n) = F(f))\}$. Then $G$ will
be the modulus for $F  = \lim_{n \rightarrow \infty}F_n.$

 Clearly $h_{n}$ and $G$ are computable as functions of $n$, $f$ and
 $F$, so we let $M_{k}(f,F) = G(f)$ as defined above.\qed

\begin{lem}\label{lemma2.3} If $a \in Ct_{\N}(k)$ and $n,m \in \N$,
then 
$((a)_n)_m = a_{min\{n,m\}}.$
\end{lem} This is proved by a trivial induction on $k$.

 Let $X^k_n = \{a_n\;;\;a \in Ct_\N(k)\}$.
\begin{lem}\label{lemma2.4} Each $X^k_n$ is a finite set. 
\end{lem}

\proof

 We use induction on $k$:

 $X^0_n = \{0 , \ldots , n\}$. 

 If $x_1 = (f_1)_n$,
$x_2 = (f_2)_n$ and $x_1(y) = x_2(y)$ for all $y 
\in X^{k-1}_n$, then, using Lemma \ref{lemma2.3}, we have for all $\xi 
\in Ct_\N(k-1)$ $$
x_1(\xi) = (x_1)_n(\xi) = (x_1(\xi_n))_n =( x_2(\xi_n))_n =
(x_2)_n(\xi) = 
x_2(\xi),$$
so $x_1 = x_2$. Since $(f)_{n}$ is bounded by $n$, we have an embedding of $X^k_{n}$ into the finite set $X_{n}^{k-1} \rightarrow \{0, \ldots , n\}$. This embedding is actually onto.\qed

 The definition of the $n$'th approximation makes
perfect sense for $a \in 
N(k)$ as well, with $(\bot)_n = \bot$. We then have
\begin{lem} \label{lem.fin} Let $a \in N(k)$ be compact.

 Then there is an $n_a \in \N$ such that for $n \geq n_a$ we
have that $a 
\sqsubseteq (a)_n$.
\end{lem}

\proof

 We use induction on $k$. For $k = 0$
this is trivial, so let $a$ be a 
compact element of $N(k+1)$.

 Then there are compact elements $b_1 , \ldots , b_r$ in
$N(k)$ and 
numbers $m_1 , \ldots , m_r$ such that $a$ is minimal with the
property 
that $a(b_i) = m_i$ for $i = 1 , \ldots , r$.

 Let $n_a = max\{n_{b_i},m_i\;;\;i \leq r\}$.

 $n_a$
will be the maximal value found in $a$ or any of 
the hereditary sub-elements of $a$.\qed

\subsection{The construction} We will now construct the maps $\pi_k$ by recursion on $k$. For $k = 0$ and $k = 1$ we will give explicit definitions. For $k > 1$ we will assume that $\pi_{{k-2}}$ is defined and satisfies the requirements of the theorem.

For $k > 1$ the definition of $\pi_{k}$ will for the sake of convenience be restricted to $\bar N(k)$. The definition is split into two cases, Case 1 and Case 2. In Case 2, the construction is easily extended to $N(k)$, just interpreting the algorithm given over the partial objects as well. The construction in Case 2 will be effective. An important part of the proof will be to show that the $\pi_{k}$ will be continuous on $\bar N(k)$.  What we really do in this argument is to extend the part of $\pi_{k}$ that is defined under Case 1 in an effective way to a partial continuous object consistent with the part of $\pi_{k}$ constructed under Case 2. In proving consistency we rely on the fact that the total objects are dense for each space under consideration. The join of these two (the constructed part under Case 2 and the extension of the part constructed under Case 1) will finally form our $\pi_{k}$.

Discussing the consequences of the theorem, we pointed out that each $\pi_{k}$ will have a partial inverse $\pi^{-1}_{k}$. Our separation in the two cases is needed in order to handle the problem that there is no total such inverse, due to the different topological nature of the Kleene-Kreisel functionals and the $Ct^{E}_{\R}$-hierarchy, where each space is path connected. In Case 1 we will use the partial inverse that will exist in this case, and in Case 2 we want in a continuous way to bridge the gaps in the construction under Case 1.

 In order to avoid too much notation, we will occasionally
view $X^k_n$ as 
a subset of $\bar N(k)$ instead of $Ct_\N(k)$.

 Let $\pi_0(n) = n$ seen as an element of $E(0)$.

 Instead of proving the induction step, we prove a slightly stronger statement that we will need in Section \ref{sec5}.
\begin{lem} \label{lemma.150405}
Let $k \geq 1$ and assume that $\pi_{0}, \ldots , \pi_{k-1}$ are constructed according to the specifications of Theorem \ref{Th.emb}.

Then there is a continuous and total $$\Pi_{k}:(N(k-1) \rightarrow E(0)) \rightarrow E(k)$$
such that $ f =  \Pi_{k}(f) \circ \pi_{k-1}$ whenever $f:N(k-1) \rightarrow E(0)$ is total.
\end{lem}
 We will obtain $\pi_{k}(f)$ for $f \in N(k)$ by first modifying $f$ to a $g:N(k-1) \rightarrow E(0)$ via the inclusion $N(0) \rightarrow E(0)$, and then use $\Pi_{k}$. It is of course sufficient to show that $\Pi_{k}$ is continuous.

\Proof {\em of Lemma \ref{lemma.150405}.\/}\
There will be one direct construction for
$k = 
1$, and one depending on $\pi_{k-2}$  for $k > 1$.

 Let $\Pi_{1}:(\N_{\bot} \rightarrow E(0)) \times E(0) \rightarrow E(0)$ be continuous such that for total $f \in \N_{\bot} \rightarrow E(0)$ and total $x \in E(0)$ we have
\begin{itemize}
\item $\Pi_{1}(f)(x) = f(0)$ if $x \leq 0$.
\item $\Pi_{1}(f)(x) = (1-y)f(n) + yf(n+1)$ when $x = n+y$ and $0 \leq y \leq
1$.
\end{itemize} 
We may choose $\Pi_{1}$ to be definable in $Real PCF$.
 From now on, let $k \geq 2$, let $F$ be a total
map in $N(k-1) \rightarrow 
E(0)$ and let $g \in \bar E(k-1)$. We will define $\Pi_{k}(F)(g) \in E(0)$ and
prove 
that $\Pi_{k}$ is continuous.

 We will use $\xi$ and $\eta$ for elements in $\bar N(k-2)$.
Let $\{\eta_n\}_{n \in 
\N}$ be an effectively enumerated dense subset of $\bar N(k-2)$.

 We will separate the definition of $\Pi_{k}(F)(g)$ into two cases, and
prove 
continuity later.

Let $\N_{E}$ be the set of elements in $E(0)$ representing natural numbers and let $nat:E(0) \rightarrow N(0)$ send representatives of $n$ to $n$.

 {\em Case 1:\/}\ $g(\pi_{k-2}(\eta_n)) \in
\N_{E}$ for all $n$.

 By continuity and the totality of $g$ we have
that $g(\pi_{k-2}(\xi)) \in 
\N_{E}$ for all $\xi \in \bar N(k-2)$. Let 
$$
f_g = \lambda \xi \in N(k-2).nat(g(\pi_{k-2}(\xi))).$$
Then $f_g \in \bar N(k-1)$ and we let $\Pi_{k}(F)(g) = \pi_{0}(F(f_g))$.

 It is at this point that we  ensure that
$\pi_k(F)(\pi_{k-1}(f)) = \pi_{0}(F(f))$, see the end of the proof.

 {\em Case 2:\/}\  Otherwise.

In order to save notation and making the construction more transparent, we behave as if we operate over $\R$ and with $\N \subseteq \R$. As mentioned above, if we view this definition as an algorithm for exact computations over the partial reals, we actually define an effective map $\Pi'_{k}\in (N(k-1) \rightarrow E(0)) \rightarrow (E(k-1) \rightarrow E(0))$ in this case, and we will let $\Pi'_{k} \sqsubseteq \Pi_{k}$ in the end.

Let $d(g,n)$ be the distance from
$g(\pi_{k-2}(\eta_n))$ to $\N$.  Since $g$ is continuous, there is an
$\epsilon > 0$ and infinitely many $n$ such that $d(g,n)>\epsilon.$
Thus $$
\sum_{n = 0}^\infty d(g,n) = \infty.$$
Let 
\begin{itemize}
\item $z_n(g) = 1$ if $\sum_{i \leq n}d(g,i) \leq 1$
\item $z_n(g) = 0$ if $\sum_{i < n}d(g,i) > 1$.
\item $z_n(g) = y$ such that $\sum_{i < n}d(g,i) + y = 1$ otherwise.
\end{itemize} Since $g$ is total, this makes sense.

 Each $x \in \R$ will induce a probability distribution $\mu_x$ on $\N$ by
\begin{itemize}
\item If $x \leq 0$, then $\mu_x(0) = 1$ and $\mu_x(n) = 0$ for $n >
0$.
\item If $n - \frac{1}{3} \leq x \leq n + \frac{1}{3}$, then $\mu_x(n)
= 1$ 
and $\mu_x(m) = 0$ for $m \neq n$.
\item If $n + \frac{1}{3} \leq x \leq n + \frac{2}{3}$, let $y \in
[0,1]$ be such 
that $x = n + \frac{1+y}{3}$.

 Then let $\mu_x(n) = 1-y$, $\mu_x(n+1) = y$ and $\mu_x(m) =
0$ for all 
other $m$.
\end{itemize}  $\mu_x$  will induce a
probability distribution 
on $X^{k-1}_n$ as follows:

 For $a \in X^{k-1}_n$, let $$
\mu_{n,g}(a) = \prod_{b \in 
X^{k-2}_n}\mu_{min\{n,g(\pi_{k-2}(b))\}}(a(b)).$$
{\em Claim 1:\/} $$
\sum_{a \in X^{k-1}_n}\mu_{n,g}(a) = 1.$$

\proof

 Each function $h:X^{k-2}_n \rightarrow \{0 ,
\ldots , n\}$ corresponds to 
one and only one $a \in X^{k-1}_n$.

 $\mu_{min\{n,g(\pi_{k-2}(b))\}}$ is a probability distribution on
$\{0 , 
\ldots , n\}$, so $\mu_{n,g}$ can be viewed as the product distribution.
Claim 1 follows.

 Now, let $$
\Pi_{k}(F)(g) = \sum_{n \in \N}\left ( d(g,n)\cdot z_n(g) \cdot \sum_{a \in 
X^{k-1}_n}(F(a) \cdot \mu_{n,g}(a))\right ).$$
This ends the construction in Case 2.

 It is easy to see that the constructions in Cases
1 and 2 are continuous 
separately. Moreover, the domain for Case 2 is open. In order to prove
the 
continuity of $\Pi_{k}(F)$ and of the map $\Pi_{k}$ it is sufficient to
show 
that if $g \in \bar R(k-1)$ falls under Case 1 and $\epsilon > 0$ is
given, 
there are compact approximations $\delta$ and $\tau$ to $F$ and $g$
resp. 
such that for any total $F' \in N(k-1) \rightarrow E(0)$ extending 
$\delta$ and any total $g' \in E(k-1)$ extending $\tau$ we have that
$$
|\Pi_{k}(F')(g') - F(g)| < \epsilon.$$
So, let $F$, $g$ and $\epsilon > 0$ be given as above, and without
loss of 
generality, assume that $\epsilon < 1$.

 Let $f \in N(k-1)$ be defined by $$
f(\xi) = nat(g(\pi_{k-2}(\xi)))$$
{\em Claim 2:\/}\
 There are $\sigma
\sqsubseteq f$, $\delta \sqsubseteq F$ and $n_0 \in \N$ 
such that $\delta(\sigma)$ has length $< \frac{\epsilon}{3}$, such
that 
for all $n \geq n_0$ and $f'$ extending $\sigma$ we have that $\sigma 
\sqsubseteq f'_n$, such that $\sigma(\tau) < n_0$ whenever defined and
such 
that if $n \leq n_0$ and $a \in X^{k-1}_n$ then $\delta(a)$ has length
$\leq \frac{\epsilon}{3}$.

\proof

 First pick $\sigma
\sqsubseteq f$ and $\delta' \sqsubseteq F$ such that 
$\delta'(\sigma)$ has length $< \frac{\epsilon}{3}$.

 By Lemma \ref{lem.fin} there is an $n_0$ such that
$(\sigma)_n = \sigma$ 
for $n \geq n_0$ and such that $n_0$ exceeds all values of $\sigma$.

 Given $n_0$, we may find $\delta \sqsubseteq F$ with $\delta
' \sqsubseteq 
\delta$ such that $\delta(a)$ has length $\leq \frac{\epsilon}{3}$ for
all $a \in X^{k-1}_n$ with $n \leq n_0$. 

 This ends the proof
of the claim.

 Let $\hat \sigma = \sigma \pm \frac{1}{3}$, i.e.
$\sigma$ is the 
compact in $N(k-2) \rightarrow E(0)$ where each value $c \in \N$ is 
replaced by the interval $[c - \frac{1}{3},c+\frac{1}{3}]$. Thus $\hat
\sigma
\sqsubseteq \lambda \xi \in N(k-2).g(\pi_{k-2}(\xi))$. Let $\tau_0 
\sqsubseteq g$ be compact such that $$
\hat \sigma \sqsubseteq \lambda \xi \in
N(k-2).\tau_0(\pi_{k-2}(\xi)).$$
Let $$
M = max\{|x| \;;\; x \in \delta(a) \wedge a \in X^{k-1}_n \wedge n
\leq  n_0\}.$$
Let $\tau_1 \sqsubseteq g$ be such that for any total $g'$ extending 
$\tau_1$ we have that $$
\sum_{n \leq n_0}d(g',n) < \frac{\epsilon}{3M}.$$
We may let $\tau_0 \sqsubseteq \tau_1$.

 We complete the proof of the continuity by showing

\noindent{\em Claim 3:\/}\
 If $F'$ and $g'$ are total
extensions of $\delta$ and $\tau_1$ resp., 
then $$
|\Pi_{k}(F')(g') - F(f)| < \epsilon.$$

\proof

 The proof will be divided
into the same cases as the construction.

 {\em Case 1:\/}\ $g'(\pi_{k-2}(\xi)) \in \N_{E}$ for all $\xi \in
\bar N(k-2)$.

 Then let $f' = \lambda \xi \in N(k-2).nat(g'(\pi_{k-2}(\xi))).$

 By the choice of $\tau_0 \sqsubseteq \tau_1$ we have that
$\sigma 
\sqsubseteq f'$. Since $\delta(\sigma)$ has length $<
\frac{\epsilon}{3}$ 
and since $\Pi_{k}(F')(g') = F'(f')$ we actually have that $|\Pi_{k}(F')(g') - F(f)| < 
\frac{\epsilon}{3}$.

 {\em Case 2:\/}\ Otherwise.

 Let $n_1$ be minimal such that $$
\sum_{n \leq n_1}d(g',n) \geq 1.$$

 Since $\sum_{n \leq n_0}d(g',n) < \frac{\epsilon}{3M}$,
$\epsilon < 1$ 
and $M \geq 1$, we see that $n_0 < n_1$. Then $$
\Pi_{k}(F')(g') = \sum_{n \leq n_0}\left (d(g',n)\cdot \sum_{a \in X^{k-1}_n}(F'(a)
\cdot
 \mu_{n,g'}(a))\right )$$
$$
+ \sum_{n_0 < n \leq n_1}\left (d(g',n)\cdot  z_n(g') \cdot
\sum_{a \in X^{k-1}_n}(F'(a) \cdot \mu_{n,g'}(a))\right ).$$
Since $\sum_{n \leq n_0}d(g',n) < \frac{\epsilon}{3M}$ and $|F'(a)|
\leq M$ 
whenever $n \leq n_0$ and $a \in X^{k-1}_n$, the first part will be 
bounded by $\frac{\epsilon}{3}$.

\noindent {\em Subclaim 3.1:\/}\ 
 If $n > n_0$, then $$
\mu_{n,g'}(\{a \in X^{k-1}_n\;;\;|F'(a) - F(f)| <
\frac{\epsilon}{3}\}) = 1.$$

\Proof{\em of subclaim:\/}\
Let $n > n_0$ and $a \in X^{k-1}_n$. We will show that we either have that $\mu_{n,g'}(a) = 0$ or that $|F'(a) - F(f)| < \frac{\epsilon}{3}$.

{\em Subcase 1.}\ For some $b \in X^{k-2}_{n}$ we have that  $\sigma(b)$ is defined and $\sigma(b) \neq a(b)$.

Then
$\sigma(b) < n_0$ 
by choice of $n_0$. Since $g'$ extends $\tau_0$, we have that 
$$
g'(\pi_{k-1}(b)) \in [\sigma(b) - \frac{1}{3},\sigma(b) +
\frac{1}{3}], $$
so the distance from $a(b)$ to $min\{n_0, g'(\pi_{k-2}(b))\}$ is at
least 
$\frac{2}{3}$. It follows that $\mu_{n,g'}(a) = 0$.

 {\em Subcase 2.}\ Otherwise.

Then $\sigma(b) = a(b)$ whenever $\sigma(b) \in
\N$ and we may extend $\sigma$ 
to $\sigma_1$ such that $\sigma_1(b) = a(b)$ for all $b \in X^{k-2}_n$
and $\sigma_1(\xi) \leq n$ for all $\xi$. Let $f'$ be a total
extension of $\sigma_1$ such that $f'(\xi) \leq n$ for all total
$\xi$.  Then by Lemma
\ref{lemma2.3} $$
f'_n(\xi) = f'(\xi_n) = \sigma_1(\xi_n) = a(\xi_n) = a_n(\xi) =
a(\xi)$$
so $f'_n = a$. 

 Thus $a = f'_n$ for some $f'$ extending
$\sigma$, and by the choice of 
$\sigma$ and $n_0$, we have that $\sigma \sqsubseteq a$. Since $\sigma
\sqsubseteq f$, $\delta(\sigma)$ has length $ < \frac{\epsilon}{3}$
and 
$\delta \sqsubseteq F \sqcap F'$ it follows that $$
|F'(a) - F(f)| < \frac{\epsilon}{3}.$$
This ends the proof of the subclaim. Continuity of the construction
will 
then follow from

\noindent{\em Subclaim 3.2:\/} $$
\left|\sum_{n_0 < n \leq n_1}\left (d(g',n)\cdot z_n(g') \cdot \sum_{a \in 
X^{k-1}_n}(F'(a) \cdot \mu_{n,g'}(a))\right ) - F(f)\right| < \frac{2\epsilon}{3}.$$

\Proof{\em of subclaim:\/}\ 
 Since $\tau_1 \sqsubseteq g'$, we have that $$
\sum_{n_0 < n \leq n_1}d(g',n) \cdot z_n(g') >
1-\frac{\epsilon}{3M}.$$
It follows that $$
\left|\sum_{n_0 < n \leq n_1}\left (d(g',n)\cdot z_n(g') \cdot \sum_{a \in 
X^{k-1}_n}(F'(a) \cdot \mu_{n,g'}(a))\right ) - F(f)\right|\leq$$
$$
\frac{\epsilon}{3M}\cdot|F(f)| + \sum_{n_0 < n \leq n_1}\left (d(g',n) \cdot 
z_n(g')\cdot \sum_{a \in X^{k-1}_n}(|F'(a) - F(f)|\cdot\mu_{n,g'}(a))\right ).$$
We have that $|F(f)| \leq M$, and by subclaim 3.1 the above is bounded
by $$
\frac{\epsilon}{3} + \sum_{n_0 < n \leq n_1}d(g',n) \cdot z_n(g')
\cdot 
\frac{\epsilon}{3} \leq \frac{\epsilon}{3} + \frac{\epsilon}{3} = 
\frac{2\epsilon}{3}.$$
This ends the proof of subclaim 3.2, of claim 3 and of the continuity
of 
the construction. Thus Lemma \ref{lemma.150405} is proved.\qed

 We may now end the proof of Theorem \ref{Th.emb}, where Lemma \ref{lemma.150405} provides us with the induction step.
Let $F_0 \in \bar N(k)$. Let $F$ be the corresponding total $F \in
N(k-1) \rightarrow E(0)$ and let $\pi_k(F_{0}) = \Pi_{k}(F)$.

 We show that $\pi_k(F)(\pi_{k-1}(f)) = \pi_{0}(F(f))$ by
induction on $k$. For $k = 1$ this is trivial. For $k > 1$, let $f \in
\bar N(k-1)$ and let $g =
\pi_{k-1}(f)$.

 By the induction hypothesis, the $f_g$ constructed from $g$
in Case 1 will be the $f$ given. Then $\Pi_{k}(F)(g) =\pi_{0}( F(f))$, i.e. $$
\pi_k(F)(\pi_{k-1}(f)) = \pi_{0}(F(f)).$$
In the proof of the continuity we started with a total $F$, a total $g$ and some $\epsilon > 0$ and showed the existence of approximations demonstrating the continuity of $\Pi_{k}$. Now the set of triples $(\delta, \tau, \epsilon)$ that are constructed in this proof will be decidable, and the part of $\Pi_{k}$ that can be constructed from this set of triples will be effective. Thus $\Pi_{k}$ will be the join of two effective partial functionals, and thus it will itself be effective. 
\begin{rem}{Note that the construction under Case 2 will not terminate for $g$ that falls under Case 1. We use this to observe that if $F \in N(k)$ is not total, we may let $f \in \bar N(k-1)$ be such that $F(f) = \bot$. Then $\pi_{k}(F)(\pi_{k-1}(f))$ will be undefined, because neither Case 1 nor Case 2 will provide us with a value. We use this to prove Corollary \ref{cor.tot}}\end{rem}

\section{The intensional functionals}\label{sec4}  An alternative approach to
higher type objects over $\R$ is based on  representations of the
reals using intensional objects.  A similar hierarchy was studied in Bauer,
Escard\'{o} 
and Simpson \cite{BES}. We gave a full treatment in Normann
\cite{Dag.Hier}. We call this hierarchy the {\em $I$-hierarchy}, where $I$ stands for `intensional'.

\begin{defi}{Let $ab_1b_2\cdots$ be an element in $\Z
\times \{-1,0,1\}^{\N^+}$, which we view as a set of functions defined
on $\N$. 

 Let $$
\rho^I_0(ab_1  b_2  \cdots ) = a + \sum_{n > 0}b_n \cdot 2^{-n}.$$
Let $Ct^I_\R(0) = \R$.}
\end{defi}
\begin{defi}{Let $I(0)$ be the algebraic domain consisting
of the empty sequence $e$, 
all finite sequences $ab_1 \ldots b_{n}$ and all infinite sequences 
$ab_1b_2\cdots$, where in the two latter cases $a \in \Z$ and
each 
$b_i \in \{-1,0,1\}$.

 $I(0)$ is ordered by sequence end-extensions.

 As we have seen, each maximal element in $I(0)$ will
determine a real via 
$\rho^I_0$. We let $\bar I(0)$ be the set of maximal elements in $I(0)$, and we let $$
ab_1b_2\cdots \approx^I_0 cd_1d_2\cdots $$
if they represent the same real.}
\end{defi}
We now extend these concepts to higher types in analogy with the constructions for the $N$-hierarchy and the $E$-hierarchy:
\begin{defi}{By recursion on $k$ we let
\begin{itemize}
\item[a)]$I(k+1) = I(k) \rightarrow I(0)$ in
the 
category of algebraic domains.
\item[b)]If $x_1$ and $x_2$ are in $I(k+1)$,
we 
let $x_1 \approx^I_{k+1} x_2$ if for all $y_1,y_2 \in I(k)$, if $y_1 
\approx^I_k y_2$, then $x_1(y_1) \approx^I_0 x_2(y_2).$
\item[c)]Let $\bar I(k+1) = \{x \in
I(k+1)\;;\;x 
\approx_{k+1}x\}$.
\item[d)]Let $\rho^I_{k+1}$ map an element
$x$ 
of $\bar I(k+1)$ to a function $\rho^I_{k+1}(x):Ct^I_\R(k) \rightarrow
\R$ 
defined as follows:$$
\rho^I_{k+1}(x)(\rho^I_k(y)) = \rho^I_0(x(y)).$$
Let $Ct^I_\R(k+1) = \{\rho^I_{k+1}(x)\;;\;x \in \bar I(k+1)\}$.
\end{itemize}}
\end{defi} One motivation for using the $I$-hierarchy is that
whenever $A \subseteq \N^\N$ and $f:A \rightarrow \R$ is continuous,
then there is an $\hat f:A \rightarrow
\bar I(0)$ such that $f(x) = \rho^I_0(\hat f(x))$ for each $x \in A$.
\begin{defi}{Let $T$ be a topological space.

 Let ${\mathcal R}(T)$ be the subtopology where the open sets are
\begin{center} $\{f^{-1}(O)\;;\;f:T \rightarrow \R$ is continuous and
$O \subseteq \R$ is open $\}$.
\end{center}}
\end{defi} Clearly every clopen set in $T$ will be
clopen in ${\mathcal R}(T)$, i.e. ${\mathcal Z}(T) \subseteq {\mathcal R}(T)$.

The following is essentially observed by Bauer, Escard\'{o} and
Simpson \cite{BES}:
\begin{prop}\label{prop.sis} Let $T$ be a topological space that is hereditarily Lindel\"{o}f.

Then the following are equivalent:
\begin{itemize}
\item[a)]For every continuous $f:T
\rightarrow \R$ there is a continuous $\hat f:T \rightarrow \bar I(0)$
such that $f(x) =
\rho^I_0(\hat f(x))$ for all $x \in T$.
\item[b)] ${\mathcal R}(T)$ is zero-dimensional
(i.e. has a basis of clopen sets).
\end{itemize}
\end{prop} 
The key problem under discussion is:
\begin{center}Is $Ct^E_\R(k) = Ct^I_\R(k)$ for a given $k$?
\end{center} For $k \leq 2$, the equality was proved in \cite{BES}. They also proved that the statement that $$Ct^{E}_{\R}(3) = Ct^{I}_{\R}(3)$$
followed from the assumption that $Ct_{\N}(2)$  is zero-dimensional.

 We will extend these results. As a tool, we will use the $S$-hierarchy introduced in  Normann \cite{Dag.Hier}. The $S$-hierarchy is not as
natural as the $E$-hierarchy and  the $I$-hierarchy, but it too is
equipped with hereditarily total  elements $\bar S(k)$, and the
extensional collapses $Ct^S_\R(k)$. The $S$-hierarchy is in some sense a {\em smoothened} $I$-hierarchy. We will  give the technical
definitions below. 

\begin{defi}{By recursion on $k$ we define the domain $S(k)$
and  the binary relation $\sim_k$ on $S(k)$ as follows 
\begin{itemize}
\item $S(0) = I(0)$ with the same ordering.

\noindent $x_1 \sim_0 x_2$ if there are maximal extensions $y_1$ of
$x_1$ and $y_2$ 
of $x_2$ such that $\rho^I_0(y_1) = \rho^I_0(y_2)$.
\item $S(k+1)$ is the set of Scott-continuous functions $f:S(k)
\rightarrow 
S(0)$ such that $$
x_1 \sim_k x_2 \Rightarrow f(x_1) \sim_0 f(x_2).$$
$S(k+1)$ is ordered by the pointwise ordering.

\noindent If $f_{1} \in S(k+1)$ and $f_{2} \in S(k+1)$, we let $f_1 \sim_{k+1} f_2$ if $$\forall x_{1} \in S(k) \forall x_{2} \in S(k)(x_1 \sim_k x_2 \Rightarrow f_1(x_1)
\sim_0 
f_2(x_2)).$$
\end{itemize}}
\end{defi}
By construction, $S(0)$ is an algebraic domain. In Normann \cite{Dag.Hier} it is proved that $S(k+1)$ is a closed subset of $S(k) \rightarrow S(0)$ and,  with the restricted ordering, is an algebraic domain. The compact objects in the sense of $S(k+1)$ are the compact objects in the sense of $S(k) \rightarrow S(0)$ that are in $S(k+1)$, but the boundedness relation is in general not the same.

The relation $\sim_{k}$ is reflexive and symmetric, but not transitive.
\begin{defi}{We let $\bar S(0)$ be the maximal elements with
the partial equivalence 
relation $\approx^S_0$ which will be $\sim_0$ restricted to the
maximal 
objects. We let $\rho^S_0 = \rho^I_0$. 

 By recursion on $k$ we then define a partial equivalence relation $\approx^S_k$ on $S(k)$ for 
each $k$ in analogy with our previous constructions of hierarchies. Let $\bar S(k) = \{x \in S(k)\;;\;x \approx^S_k x\}$, and
define 
$\rho^S_k$ in analogy with $\rho^I_k$.

We then define $Ct^S_{\R}(k)$ in analogy with $Ct^{I}_{\R}(k)$.}
\end{defi}

 \begin{prop}\label{prop4.8}{\em (Normann \cite{Dag.Hier})}
\begin{itemize} \item[a)] Uniformly in
any compact element $p$ in $S(k)$ there is an extension to an 
element $\xi(p) \in \bar S(k)$. \item[b)] For
each $k$, $Ct^S_\R(k) = Ct^I_\R(k).$ \end{itemize}
\end{prop} 
\begin{prop}{\em(Normann \cite{Dag.Hier})}\label{Prop.5} If $x_1,x_2
\in \bar S(k)$, then $$
x_1 \approx^S_k x_2 \Leftrightarrow x_1 
\sim_{k}¥ x_2.$$
Moreover, $\sim_k$ is a closed relation on $S(k)$.

 Finally, if $y_1$ and $y_2$ are compacts in $S(k)$ such that
$y_1 \sim_k y_2$, then there are total extensions $x_1$ and $x_2$ of $y_1$
and $y_2$ resp. such that $x_1 \approx^S_k x_2$.
\end{prop} 
Our aim is to show that the hierarchies $\{Ct^{I}_{\R}(k)\}_{k \in \N }$ and $\{Ct^{E}_{\R}(k)\}_{k \in \N}$ are identical, assuming that ${\mathcal R}(Ct_{\N}(k))$ is zero-dimensional for each $k$. In order to make use of this assumption, we will consider the $S$-hierarchy and the quotient space of $\bar S(k)$ under the consistency relation. This will be an intermediate stage between $\bar S(k)$ and $\bar S(k)/\approx_{k}^{S}$, an intermediate stage that will enjoy some of the topological qualities of $Ct_{\N}(k)$. (See Lemma \ref{lemma.key} below.)
\begin{defi}{Let $x_1,x_2 \in \bar S(k)$.  
 We
define the relation $C_k$ on $\bar S(k)$ by:

 $C_k(x_1,x_2) \Leftrightarrow x_1$ and $x_2$ are consistent,
i.e. bounded 
in $S(k)$.}
\end{defi}
\begin{lem} $C_k$ is an equivalence relation on $\bar
S(k)$.\end{lem}

\proof

 $C_0$ is the identity relation
on $\bar S(0)$.

 By Proposition \ref{prop4.8} a), two higher type
total objects in the $S$-hierarchy are 
consistent if and only if they are identical when restricted to total 
inputs. This defines an equivalence relation.\qed
\begin{defi}\label{definition.t}{Let $T(k) = \bar S(k)/C_k$ with the quotient
topology.}\end{defi}
Recall that the topology on $T(k)$ then will be sequential.
\begin{lem}\label{lemma.250405}\hfill
\begin{itemize}
\item[{a)}]
$T(0)$ is homeomorphic to $\Z \times \{-1,0,1\}^{\N^+}$.
\item[{b)}] $T(k+1)$ is homeomorphic to a closed subspace of $T(k) \rightarrow \Z \times \{-1,0,1\}^{\N^+}$.\end{itemize}
\end{lem}

\proof
a) is trivial. In order to prove b), we observe that $\bar S(k+1)$ will consist of the total elements in $S(k) \rightarrow S(0)$ that send extensionally equivalent elements in $\bar S(k)$ to extensionally equivalent elements of $\bar S(0)$.

Being extensionally equivalent is a closed relation on $\bar S(0)$, so $\bar S(k+1)$ will be a closed subset of the set of total elements from $\bar S(k)$ to $\bar S(0)$.

$C_{k+1}$ is just the restriction of the equivalence relation of consistency for total elements on $S(k) \rightarrow S(0)$, and $\bar S(k+1)$ will consist of full equivalence classes for this consistency relation.

We then obtain the lemma from Lemma \ref{lem.gen}.\qed

\begin{lem}\label{lemma.key} If $k \geq 1$, then $T(k)$ is
homeomorphic to a $ \tilde \Pi^0_1$-subspace of 
$Ct_\N(k)$.
\end{lem}

\proof

 For $k = 1$, this is proved in
\cite{BES}. Our proof is inspired by the 
proof in \cite{BES}, but we give a slightly different proof in order to
prepare 
for the induction step.

 If $f \in T(1)$, then $f$ is (the  equivalence
class representing) a total 
map from $\Z \times \{-1,0,1\}^{\N^+}$ to  $\Z \times
\{-1,0,1\}^{\N^+}$.

 $\Z \times \{-1,0,1\}^{\N^+}$ is homeomorphic to a
$\sigma$-compact 
subset of $\N^\N$, i.e. a countable union of compact sets.

 For $a \in \Z $ and $g \in \{-1,0,1\}^{\N^{+}}$ let $f_{a,i}(g)$ be the $i$'th element in the sequence
$f(a*g)$ (where $*$ 
is concatenation between a finite sequence and a function). Using the
fan 
functional, we may find a number $k(f,a,i)$ that codes the behavior of
$f_{a,i}$ on $\{-1,0,1\}^{\N^+}$ and such that $f_{a,i}$ is
recoverable from 
$k(f,a,i)$.

 If $\{a_{n}\}_{n \in \N}$ is a 1-1 enumeration of $\Z$, we let $c_{f}(n,i) = k(f,a_{n},i)$, and we let
$$X_{1} = \{c_{f}\;;\; f \in T(1)\},$$ then $X_{1}$ is homeomorphic to $T(1)$.

We call $c_{f}$ the {\em code for} $f$.

 In order to see that $X_1 \in \tilde \Pi^0_1$, we observe
that the 
$\tilde \Pi^0_{1}$-subsets of $Ct_\N(1)$ are exactly the closed
subsets, 
and further that
\begin{enumerate}
\item The set of codes in $\N^\N$ for total elements in  $$
\Z \times 
\{-1,0,1\}^{\N^+} \rightarrow \Z \times \{-1,0,1\}^{\N^+}$$
is closed.
\item The set of codes for total elements in $$
\Z \times 
\{-1,0,1\}^{\N^+} \rightarrow \Z \times \{-1,0,1\}^{\N^+}$$
that represents 
elements in $T(1)$ is closed.
\end{enumerate} (1) is trivial. (2) is seen as follows: Let $\{(\sigma_n
, \tau_n)\}_{n \in 
\N}$ be an enumeration of all pairs $(\sigma,\tau)$ of compact
elements in $S(0)$ such that 
$\sigma \sim_0 \tau$. By the third part of Proposition \ref{Prop.5},
let 
$\xi_n$ and $\eta_n$ be equivalent, total extensions of $\sigma_n$ and
$\tau_n$. Then $$
f:\Z \times 
\{-1,0,1\}^{\N^+} \rightarrow \Z \times \{-1,0,1\}^{\N^+}$$
represents an element  in $T(1)$ if and only if $$
\forall n (f(\xi_n) \sim_0 f(\eta_n))$$
if and only if $$
\forall n \forall i (\overline{f(\xi_n)}(i) \sim_0 
\overline{f(\eta_n)}(i),$$
where $\overline g(i) = (g(0) , \ldots , g(i-1))$ whenever $g$ is
defined on $\N$.

 The matrix defines a clopen set, so we are through with the
induction  start.

 Now assume that $T(k)$ is homeomorphic to a
$\tilde \Pi^0_1$-subset $X_k$ 
of $Ct_\N(k)$.

By Lemma \ref{lemma.250405}, $T(k+1)$ is homeomorphic to a closed subset $A$ of $T(k) \rightarrow \Z \times \{-1,0,1\}^{\N^+}$, and using the argument for 2. under the case for $k = 1$, we see that $A$ is indeed a countable intersection of clopen sets.

Clearly $T(k) \rightarrow \Z \times \{-1,0,1\}^{\N^+}$ is homeomorphic to $X_{k }\rightarrow \Z \times \{-1,0,1\}^{\N^+}$. Furthermore, $\Z \times \{-1,0,1\}^{\N^+}$ is homeomorphic to a $\tilde \Pi^0_{1}$-subset of $\N^\N$, and $X_{k} \rightarrow \N^\N$ is homeomorphic to $X_{k} \times \N \rightarrow \N$.

By Corollary \ref{cor.3.15}, we see that $X_{k} \rightarrow \Z \times \{-1,0,1\}^{\N^+}$ then is homeomorphic to a $\tilde \Pi^0_{1}$-subset of $Ct_{\N}(k+1)$.

If we use all these homeomorphisms to map $A \subseteq T(k) \rightarrow \Z \times \{-1,0,1\}^{\N^+}$ to $B \subseteq Ct_{\N}(k+1)$, we see that $B$ will be $\tilde \Pi^0_{1}$ and homeomorphic to $T(k+1)$. This ends the proof of the lemma.\qed

\begin{rem}{With some care, we may prove that the sets are
$\Pi^0_1$, i.e. they will 
be the intersection of an effectively given sequence of clopen 
sets.}
\end{rem} 
\begin{lem}\label{lemma.290805}
Let $(\Omega,\bar \Omega)$ be the lazy natural numbers as defined in the proof of Lemma \ref{lemma.210405}.

Then $\bar \Omega \times T(k)$ with the quotient topology is homeomorphic to a $\tilde \Pi^0_{1}$-subspace of $Ct_{\N}(k)$.\end{lem}

\proof

Clearly $\bar \Omega \times Ct_{\N}(k)$ is homeomorphic to a $\Pi^0_{1}$-subset of $Ct_{\N}(k)$. Then the lemma follows from Lemma \ref{lemma.key}.\qed

Now we are ready to prove
\begin{thm}\label{theorem.300805} If $Ct_\N(n)$ is zero-dimensional, then
$$
Ct_\R^S(n+1) = 
Ct_\R^E(n+1)$$ as topological spaces.
\end{thm}

\proof

 By the assumption, $Ct_{\N}(k)$ is zero-dimensional for $k \leq n$, i.e. the assumption of the theorem holds for all $k \leq n$.

By a simultaneous induction on $n$ satisfying the assumption, we will prove the following three claims:
\begin{enumerate}
\item $Ct^S_{\R}(n+1) = Ct^{E}_{\R}(n+1)$ as sets.
\item The quotient topology on $Ct^S_{\R}(n+1)$ coincides with the quotient topology on $Ct^{E}_{\R}(n+1)$.
\item
The quotient topology on $\bar \Omega \times Ct^S_{\R}(n+1)$ coincides with the quotient topology on $\bar \Omega \times Ct^{E}_{\R}(n+1)$.
\end{enumerate}
For $n = -1$ the three claims will hold, so we are on safe ground in proving this by inducton. Let $n \geq 1$ and assume that the three claims hold for $m = n-1$.

\Proof{\em of (1).\/}\ As remarked in Section \ref{sec2},
$Ct^{E}_{\R}(n+1)$ consists of exactly all continuous functions
$F:Ct^{E}_{\R}(n) \rightarrow \R$. It follows from the induction
hypothesis, (1) and (2), that $Ct^S_{\R}(n+1) \subseteq
Ct^{E}_{\R}(n+1)$.

By Lemma \ref{lemma.key} and the assumption it follows that $T(n)$ is
zero-dimensional, and in particular, ${\mathcal R}(T(n))$ is zero-dimensional. 
Thus, if $f \in Ct^{E}_{\R}(n+1)$, then $f:Ct^S_{\R}(n) \rightarrow \R$ is continuous, and it factors through a continuous $\hat f:T(n) \rightarrow \R$. It follows from Proposition \ref{prop.sis} that $f \in Ct^S_{\R}$, and the equality of the two sets is established.

\Proof{\em of (2).\/}\
Since both $Ct^S_{\R}(n+1)$ and $Ct^{E}_{\R}(n+1)$ are sequential topological spaces, it is sufficient to show that the convergents sequences are the same for the two topologies.

This amounts to prove that $\bar \Omega \times Ct^{E}_{\R}(n) \rightarrow \R$ and $\bar \Omega \times Ct^S_{\R}(n) \rightarrow \R$ are identical as sets, where all topologies are the relevant quotient topologies.

Using Lemma \ref{lemma.290805} instead of Lemma \ref{lemma.key}, we may use the same argument as in (1).

\Proof{\em of (3).\/}\ 
Following the line of thought from (2), we have to prove that $\bar \Omega \times Ct^{E}_{\R}(n) \rightarrow \R$ and $\bar \Omega \times Ct^S_{\R}(n) \rightarrow \R$ as topological spaces have the same convergent sequences. But in the quotient topologies of these products, the convergent sequences will be exactly the products of convergent sequences in the factors (using liftings of convergent sequences to see that the product of two convergent sequences is convergent), and then (3) follows from (2).\qed

 We also obtain the following:
\begin{thm}\label{Theorem.eq} Assume that ${\mathcal R}(Ct_\N(n))$ is
zero-dimensional. Then  $$
Ct_\R^S(n+1) = 
Ct_\R^E(n+1).$$
\end{thm}

\proof

First we observe that ${\mathcal R}(Ct_{\N}(k))$ is zero-dimensional for each $k \leq n$

{\em Claim:\/}\ 
If $k \leq n$ and $A \subseteq Ct_{\N}(k)$ is $\tilde \Pi^0_{1}$, then ${\mathcal R}(A)$ is zero-dimensional.

 {\em Proof of claim:\/}\ 
Let $f:A \rightarrow \R$ be continuous and $O \subseteq \R$ be open.

By Theorem \ref{Theorem.ext} a), $f$ may be extended to a continuous $g:Ct_{\N}(k) \rightarrow \R$. Then, by the assumption, $g^{-1}[O]$ will be the union of clopen sets in $Ct_{\N}(k)$, and thus $f^{-1}[O]$ will be the union of clopen sets in $A$.

This ends the proof of the claim.

From now on we may use the same argument as in the proof of Theorem \ref{theorem.300805}.

Since $Ct^{S}_{\R}(k) = Ct^{I}_{\R}(k)$ with the same topology, we
have proved that if ${\mathcal R}(Ct_{\N}(k))$ is zero-dimensional for
each $k$, the extensional and intensional hierarchies will
coincide.\qed 

\section{A topological characterization}\label{sec5} We have only used Proposition
\ref{prop.sis} one way.  We
will now prove the converse of Theorem \ref{Theorem.eq}, using the other direction of Proposition \ref{prop.sis}.

As a tool, we will construct continuous maps $\pi^{S}_{k}:\bar N(k) \rightarrow \bar S(k)$ and embeddings  $\bar \pi^{S}_k:Ct_\N(k) 
\rightarrow Ct_\R^S(k)$ using as far as possible the same construction as for
$\pi_k$ and $\bar \pi_{k}$ from the proofs of Theorem \ref{Th.emb} and its corollaries. 

 In Normann \cite{Dag.Hier} we proved that the hierarchies 
$\{Ct^E_\R(k)\}_{k \in \N}$ and $\{Ct^I_\R(k)\}_{k \in \N}$ have a
maximal 
common core, i.e. there are maximal isomorphic sub-hierarchies of these
type 
structures.
Using the similarities of the constructions of $\bar \pi_{k}$ and $\bar \pi^{s}_{k}$ we obtain that the $Ct_\N$-hierarchy can be embedded
into this core, see Theorem \ref{th.hier}.
\begin{lem}\label{lemma.help} There is a total, continuous map,
called the ``normalizer", $$
norm:S(0) \rightarrow S(0)$$
representing the identity map on $\R$ such that whenever $x \in \bar
S(0)$ 
and 

 $\rho^S_0(x) = n \in \N$, then $norm(x) = n00 \cdots$ .
\end{lem}

\proof

 From finite information about $x$ we may split between two
overlapping 
cases:
\begin{enumerate}
\item $\exists n (\rho^S_0(x) \in (n - \frac{1}{2}, n + \frac{1}{2}))$
\item $\exists n (\rho^S_0(x) \in (n + \frac{1}{3}, n +
\frac{2}{3}))$.
\end{enumerate}In Case (2) we let $norm(x) = x$.

 In Case (1) we find the $k$'th element in the sequence
$norm(x)$ by 
recursion on $k$ as follows: 

 The first element is $n$ (we are
certain that $n$ has an extension 
equivalent to $x$).

 Assume that we at the $k$'th stage have decided that
$norm(x)$ starts with 
$n$ and then $k-1$ zeros, and that we know that $x$ represents a real
in 
$(n-2^{-k},n+2^{-k})$.

 Then from a finite part of $x$ we may separate between the
overlapping cases
\begin{itemize}
\item $x$ represents a real in $(n - 2^{-(k+1)},n + 2^{-(k+ 1)})$
\item $x$ represents a real in $(n - 2^{-k} , n - 2^{-(k+2)})$
\item $x$ represents a real in $(n + 2^{-(k+2)},n+2^{-k})$.
\end{itemize} In the first case, we add a new zero to $norm(x)$ and
continue. In the other 
cases, we let $norm(x)$ be an extension of $n*(0)^{k-1}$ equivalent to
$x$, which we can find continuously in $x$.\qed

\begin{thm}\label{lemma.emb.alt}For each $k \in \N$ there is a
continuous 
total map $$
 \pi^{S}_k:N(k) \rightarrow S(k)$$
such that whenever $x \in \bar N(k+1)$ and $y \in \bar N(k)$ we have
$$
\rho^N_0(x(y)) = \rho^S_0( \pi^{S}_{k+1}(x)( 
\pi^{S}_k(y))).$$
\end{thm} 
We will prove the theorem below. Let us first observe:
\begin{cor}\label{cor.new} For each $k \in \N$ there is an injective, continuous map $$\bar \pi^{S}_{k}:Ct_{\N}(k) \rightarrow Ct^{S}_{\R}$$ such that
\begin{itemize}
\item[{i)}]
$\bar \pi^{S}_{0}$ is the standard inclusion map from $\N$ to $\R$.
\item[{ii)}] For each $k \in \N$, each $F \in Ct_{\N}(k+1)$ and each $a \in Ct_{\N}(k)$ we have that
$$F(a) = \bar \pi^{S}_{k+1}(F)(\bar \pi^{S}_{k}(a)).$$
\item[{iii)}] For each $k \in \N$, the range of $\bar \pi^{S}_{k}$ is a closed subset of $Ct^{S}_{\R}(k)$ homeomorphic to $Ct_{\N}(k)$.
\end{itemize}\end{cor}

\proof

Items i) and ii) are direct consequences of Theorem \ref{lemma.emb.alt}.

Item iii) is proved like Corollary \ref{cor.180405} with the obvious adjustment to $S(0)$ of the first bullet point. The adjustment is obvious since each compact element of $S(0)$ determines a closed, rational interval.\qed

\Proof {\em of Theorem \ref{lemma.emb.alt}.\/}\ 
 We will adjust the construction of $\pi_{k}$ used
to prove Theorem \ref{Th.emb}. In order to give a sound construction we first had to define $\pi_{k}$ and prove its properties, and then define $\bar \pi_{k}$. However, the underlying way of thinking goes the other way, we define $\bar \pi_{k}$ and then show that it is continuous by constructing an appropriate $\pi_{k}$.

Our definition of $\bar \pi^{S}_{k}$ will be almost like our definition of $\bar \pi_{k}$, the difference is that now the definition is interpreted over the intensional hierarchy while in the original case, it was interpreted over the extensional hierarchy. The challenge then is to show that $\bar \pi^{S}_{k}$ can be realized by an intensional object. In order to avoid repeating the details of the proof of Theorem \ref{Th.emb}, we will only discuss the obstacles that are new in the intensional setting. 

 {\em Case $k = 0$:\/}\
 Let $\bar \pi^{S}_{0}(n) = n \in \R$ realized by $ \pi^{S}_0(n) =
n00\cdots $ .

 {\em Case $k = 1$:\/}\
 Let $f:\N \rightarrow\N$ and $x \in \R$.

 We let  $\bar \pi^{S}_{0}(f)(x) = f(0)$ if $x
\leq 0$ and 
$$
\bar \pi^{S}_{0}(f)(x) = (1-y)f(n) + yf(n+1)$$
when $x = n+y$ and $0 \leq y \leq 1$.
 Since $\N^\N \times \bar S(0)$  is zero-dimensional, it follows from 
Proposition \ref{prop.sis} that there is a total  $ 
\pi^{S}_1:N(1) \times S(0) \rightarrow S(0)$ realizing $\bar \pi^{S}_1$.

 {\em Case $k \geq 2$:\/}\
 Let $F \in Ct_{\N}(k)$ and let $x \in Ct^{S}_{\R}(k-1)$.

For $\xi \in Ct_{\N}(k-2)$, let $f(\xi) = x(\bar \pi^{S}_{k-2}(\xi)).$

 Let Case 1 and Case 2 correspond to the cases in the proof of
Theorem 
\ref{Th.emb}. In Case 1, $f:Ct_{\N}(k-2) \rightarrow \N$ and we let
$$
 \pi^{S}_k(F)(x) = F(f).$$
In  Case 2, we define $$
\bar \pi^{S}_{k}(F)(x) = \sum_{n = 0}^\infty \left (d(x,n)\cdot z_n(x)\cdot\sum_{a \in 
X^{k-1}_n}(F(a)\cdot\mu_{n,x}(a)) \right )$$ where we use the notation from the proof of Theorem \ref{Th.emb}.

$\bar \pi^{S}_{k}(F)(x)$ will depend continuously on the sequences $\{ x(\bar \pi^{S}_{k-2}(\eta_n))\}_{n \in \N}$ and
$\{F(a)\}_{n \in \N,a \in X^{k-1}_n}$.

Here we view $\eta_{n}$ as an element of $Ct_{\N}(k-2)$ and $X^{k-1}_{n}$ as a subset of $Ct_{\N}(k-1)$.

 Let $\bar G$ be continuous such that
$$\bar \pi^{S}_{k}(F)(x) =  \bar G(\{ x(\bar \pi^{S}_{k-2}(\eta_n))\}_{n \in \N},\{F(a)\}_{n \in \N,a \in X^{k-1}_n})$$ when we are in Case 2.

 $\bar G$ is essentially of type $\R^\N \rightarrow \R$ and  can be represented in the  $S$-hierarchy by a total, continuous $G$. We use $G$
 to define $ \pi^{S}_k(F)(x)$  in a continuous
way from 

 $\{x(\tilde
\pi_{k-2}(\eta_n))\}_{n \in \N}$  and $\{F(a)\}_{n \in \N,a \in
X^{k-1}_n}$.

 If we do this without any further care, we will not be able
to show that the 
constructions from Case 1 and Case 2 match in a continuous way. However, if we
use the intensional representation $F(f)00\cdots$ in Case 1, and $norm \circ G$ in Case 2 (where $norm$ is the function of Lemma \ref{lemma.help}), we may prove continuity in the
same 
way as we did in the proof of Theorem \ref{Th.emb}.

This ends our proof of the theorem.\qed

Our next result relate the two embeddings to the core hierarchy from Normann \cite{Dag.Hier}. We will not need this result elsewhere in this paper, so we assume familiarity with \cite{Dag.Hier} in this proof.
\begin{thm}\label{th.hier} Let $x \in Ct_{\N}(k)$. Then $\bar \pi_k(x) \in
Ct^E_\R(k)$ and 
$\bar \pi^{S}_k(x) \in Ct^S_\R(k)$ are equivalent in the sense
of 
Normann \cite{Dag.Hier}.
\end{thm}

\proof

 We use induction on $k$, where the
first two cases are trivial, so let $k 
\geq 2$.

 Objects from these type structures are equivalent if they
behave in the same 
way on equivalent input. We observe that $\bar \pi_k(F)(x)$ only
depends on $x$ restricted to the image of $\bar \pi_{k-2}$ and that $\bar
\pi^{S}_k(F)(x)$ depends on $x$ restricted to the image of $\bar \pi^{S}_{k-2}$ in the same
way. 
Thus if $F \in Ct_{\N}(k)$ is given and $x \in Ct_{\R}^{E}$ and $y \in Ct_{\R}^{S}$ are equivalent at type $k-1$ we use exactly the
same 
definition in the two cases, and the results will be two equivalent 
functionals.\qed

\begin{thm}Assume that $Ct_\R^E(n+1) = Ct^S_\R(n+1)$.

 Then ${\mathcal R}(Ct_\N(n))$ is zero-dimensional.
\end{thm}

\proof

First observe that by the assumption it follows, for purely set theoretical reasons, that $Ct_{\R}^{E}(k) = Ct_{\R}^{S}(k)$ for $k \leq n$.

Let $f:Ct_\N(n) \rightarrow \R$ be
continuous.

Then there is a total, continuous function $h:N(n) \rightarrow E(0)$ such that
$$\rho^{E}_{0}(h(x)) = f(\rho_{n}^{N}(x))$$
for all $x \in \bar N(n)$.

Let $\Pi_{n+1}$ be as in Lemma \ref{lemma.150405}, and let $g = \rho^{E}_{n+1}(\Pi_{n+1}(h))$.

Then for all $x \in Ct_{\N}(n)$ we have that $f(x) = g(\bar \pi_{n}(x)).$

Since $g \in Ct^{E}_{\R}(n+1)$ it follows from the assumption that $g \in Ct^{S}_{\R}(n+1)$, which means that there is a total $\hat g \in \bar S(n+1)$ with $g = \rho^{S}_{n+1}(\hat g)$.

Recall the topological space $T(n)$ from Definition \ref{definition.t}. $T(n)$ is $\bar S(n)$ divided out by the consistency relation $C_{n}$ on $S(n)$.

Clearly $\hat g$ will send consistent, total elements in $ S(n)$ to consistent, total elements in $S(0)$, and consistency on $\bar S(0)$ is the same as identity, so $\hat g$ will induce a total $\tilde g : T(n) \rightarrow \bar S(0)$.

Let $O \subseteq \R$ be open. Then
$$O^{S} = \{y \in \bar S(0)\;;\;\rho^{S}_{0}(y) \in O\}$$ is open in $\bar S(0)$, and since $\bar S(0)$ is zero-dimensional, $O^{S}$ will be the union of clopen sets. Then
$$\{z \in T(n)\;;\;\rho^{S}_{0}(\tilde g(z)) \in O\}$$
is the union of clopen sets.

Let $\pi^{S}_{n}:N(n) \rightarrow S(n)$ be as in Theorem \ref{lemma.emb.alt}.

We define the map $\pi^{T}_{n}$ as follows: Let $x \in Ct_{\N}(n)$, let $y \in \bar N(n)$ be such that $\rho^{N}_{n}(y) = x$ and let $\pi^{T}_{n}(x) \in T(n)$ be the equivalence class of $\pi^{S}_{n}(y)$ (which is independent of the choice of $y$).

Then $\pi^{T}_{n}$ is continuous, and $$f^{-1}[O] = (\pi^{T}_{n})^{-1}[\tilde g^{-1}[\{y \in S(0)\;;\;\rho^{S}_{0}(y) \in O\}]].$$
It follows that $f^{-1}[O]$ is the union of clopen sets.

Since $f$ and $O$ was arbitrary, ${\mathcal R}(Ct_{\N}(n))$ will be
zero-dimensional.\qed

\section{Discussion} \label{sec6}We have established the equivalence of a problem
about functionals over 
the reals with a problem in topology, see Problem \ref{problem2}. In fact, there will be several interesting problems of topology related to the coincidense problem:
\begin{prob}\label{problem1} Is $Ct_\N(k)$ zero-dimensional for
some or 
all $k > 1$?
\end{prob}
\begin{prob}\label{problem2} Is ${\mathcal R}(Ct_\N(k))$
zero-dimensional 
for some or all $k >1$?
\end{prob} 
\begin{prob}\label{problem3}{Does $Ct_{\N}(k)$ coincide, as a topological space, with ${\mathcal R}(Ct_{\N}(k))$ for some or all $k > 1$}?\end{prob}
Of course, if Problem \ref{problem1} has a positive
solution for a value $k$, then 
Problem \ref{problem2} will also have a positive solution for the same
$k$, and if both Problems \ref{problem2} and \ref{problem3} have positive solutions for some $k$, then Problem \ref{problem1} has a positive solution for the same $k$.

 Some effort has been put into solving Problem \ref{problem1},
but without 
success. At the time of writing, three possible answers to Problem 
\ref{problem1} seem equally likely, the answers `yes, `no' and 
`independent of $ZFC$'. There is no indication from the attempts so
far 
that Problem \ref{problem2} or Problem \ref{problem3} are easier to solve. The author sees no
basis for 
forming any conjecture.

There are topological aspects of these problems that we will not discuss in this paper, but it may well be that a deeper use of topology is what is needed to solve them. Another possibility is, as mentioned, that some or all of the problems are independent of Zermelo-Fraenkel set theory.

The statement
\begin{center}
$Ct_{\N}(2)$ is zero-dimensional \end{center}
is $\Pi^1_{4}$, and even, given an open set $O \subseteq Ct_{\N}(2)$, the statement that \begin{center}
$O$ contains a nonempty clopen set \end{center}
is $\Sigma^1_{3}$ relative to the set of compacts approximating elements in $O$. It is known that the  truth value of some $\Sigma^1_{3}$-statements can be altered by forcing. It must be stressed that so far there is nothing indicating that this will be possible in this case. The only thing we actually know is that by the Shoenfield Absoluteness Theorem, being an open subset of $Ct_{\N}(2)$ is absolute with respect to forcing, being a clopen subset of $Ct_{\N}(2)$ is absolute with respect to forcing, and thus, if a fixed open set is the union of clopen sets, this cannot be altered by forcing. Thus forcing a counterexample must mean constructing a new open set. Open sets in $Ct_{\N}(2)$ constructed by a naive use of forcing, will however contain clopen sets. The use of forcing in domain theory has not been developed sufficiently far for us to comment further on this.

\section{Appendix}\label{sec7} In this appendix we will prove a special case of
the approximation lemma from Normann  \cite{Dag.Def}:
\begin{thm} \label{th.apprx}Let $A \subseteq Ct_\N(k)$ and let $f:A \rightarrow
\N$ be continuous. Then, continuously in $f$ there are $f_n \in
Ct_\N(k+1)$ for each $n \in \N$ such that whenever $x \in A$ and $x = \lim_{n \rinf} x_n$
with each $x_n \in Ct_\N(k)$ we have that $f(x) =
\lim_{n \rinf}f_n(x_n)$.
\end{thm}

\proof

 Let $\bar A = \{x \in \bar
N(k)\;;\;\rho^N_k(x) \in A\}$. We will let $f \in N(k+1)$ be total on
$\bar A$ in the proof.  

 Let $\{(p_i,a_i)\}_{i \in \N}$ be an
enumeration of all pairs $(p,a)$ where $p
\in N(k)$ is compact, $p$ has an extension in $\bar A$ and $a \in \N$.

 If $i$ and $j$ are such that $p_i$ and $p_j$ have a joint
extension in $\bar A$, let $z_{i,j}$ be one such extension. We will
consider $i,j$ as an unordered pair. In particular, $z_{i,i}$ will
exist for all $i$.

 Let $X^f_n = \{(p_i,a_i)\;;\;i \leq n$ and for all $j \leq
n$, if $z_{i,j}$ exists, then $f(z_{i,j})=a_i\}$.

 Let $(p_j,a_j) \in Y^f_n$ if $j \leq n$ and for some $r \leq
n$ we have that 
\begin{itemize}
\item $(p_r,a_{r}) \in X^f_n$.
\item $a_j = a_r$.
\item $p_r \sqsubseteq p_j$.
\item If $i < r$ and $(p_i,a_i) \in X^f_n$, then $a_i = a_r$ or $p_i$
and $p_j$ are inconsistent. \end{itemize} It is easy to see that if
$(p_j,a_j) \in Y^f_n$, $(p_{j'},a_{j'}) \in Y^f_n$ and $p_j$ and
$p_{j'}$ are consistent, then $a_j = a_{j'}$.

 By the density theorem for $Ct_\N(k+1)$ there is a total map
$g = g_{Y^f_n}$ such that $g(p_j) = a_j$ whenever $(p_j,a_j) \in
Y^f_n$. We let $f_n = g_{Y^f_n}$.

 Now, let $x = \lim_{n \rinf}x_n$, where $x \in \bar A$ and
each $x_n \in \bar N(k)$. Let $f(x) = a$. Then there is an
approximation $p$ to $x$ such that $f(p) = a$.

 For some $r \in \N$ then, $(p,a) = (p_r,a_r)$. Moreover, if
$r \leq n$, $i \leq n$ and $z_{r,i}$ exists, then $z_{r,i}$ extends
$p$, so $f(z_{r,i}) = a = a_r$.

 Thus $(p_r,a_r) \in X^f_n$ whenever $n \geq j$.

 Let $i < r$, $n \geq r$, $(p_i,a_i) \in X^f_n$ and $a_i \neq
a_r$.

\noindent {\em Claim:\/}\
 $a_i$ and $x$ are inconsistent.

\proof

 Assume that $p_i$ and $x$ are
consistent. Then $p_i \sqcup x$ is a joint extension of $p_i$ and
$p_r$ in $\bar A$, so $z_{i,r}$ exists. Then $$
(p_i,a_i) \in X^f_n \Rightarrow f(z_{i,r}) = a_i$$
and $$
(p_r,a_r) \in X^f_n \Rightarrow f(z_{i,r}) = a_r,$$
contradicting $a_i \neq a_r$. This proves the claim.

 Then there is a compact element $p \sqsubseteq
x$ such that $p_r \sqsubseteq p$ and $p$ is inconsistent with $p_i$
whenever $i < r$ and $a_i \neq a_r$.

 For some $j$ then, $(p,a) = (p_j,a_j)$.

 If $n \geq max\{r,j\}$, it follows that $(p_j,a) \in Y^f_n$,
so $f_n(p_j) = a$.

 Let $n_0 \geq max\{r,j\}$ be such that $n \geq n_0
\Rightarrow p_j \sqsubseteq x_n$. Then $n \geq n_0 \Rightarrow
f_n(x_n) = a$, and this is what we aimed to prove.

 If $y = \lim_{n \rinf}y_n$ in $Ct_\N(k)$, there
will be $x = \lim_{n \rinf}x_n$ in $\bar N(k)$ such that $y =
\rho^N_k(x)$ and $y_n = \rho^N_k(x_n)$ (see Proposition \ref{prop.lim}). Moreover, the $f_n$
constructed only depends on $f$ restricted to $\bar A$, and is thus
definable from $\rho_A(f):A \rightarrow \N$, where $$
\rho_A(f)(\rho^N_k(x)) = f(x).$$
Actually, $f_{n}$ is definable from $f$ restricted to
\begin{center}$\{\rho^{N}_{k}(z_{i,j})\;;\; i \leq n, j \leq n$ and $z_{i,j}$ exists $\}$,\end{center} so $f_{n}$ depends continuously on $f$.
Now, by Proposition \ref{proposition.new} , every continuous function from $A$ to $\N$ will be of the form
$\rho_A(f)$, so we are through.\qed

\end{document}